\documentclass[12pt,reqno]{article}

\usepackage{latexsym}
\usepackage{amssymb}
\usepackage{euscript}
\usepackage{eufrak}
\newcommand{\sumhalf}[2]                                       {%
\renewcommand{\arraystretch}{.75}
\begin{array}{c}
\mbox{$_{#2}$} \\
\displaystyle\sum \mbox{\scriptsize{$\!\!_\frac{1}{2}$}} \\
\mbox{\large ${_{^{#1}}}$}
\end{array}
\renewcommand{\arraystretch}{1}                            } 

\begin{document}

\title{Concept of a veritable osp(1$|$2) super-triangle sum rule with $6$-$j^S$ symbols from intrinsic operator techniques: an open problem \\}
\author{L. Br\'{e}hamet \thanks {E-mail: brehamet.lionel@aliceadsl.fr \it(formerly : lionel.brehamet@cea.fr)} \\{\small CENTRAL LAMARQUE }, 74 Crs Lamarque \\ 33120 Arcachon, France}
\maketitle
\begin{abstract}
Efficiency of intrinsic operator techniques (using only products and ranks of tensor operators)
is first evidenced by condensed proofs of already known $\bigtriangledown$-triangle sum rules of 
su(2)/su$_q$(2). {\em A new compact} su$_q$(2)-{\em expression} is found, using a $q$-series
$\Phi$, with $\Phi(n)_{| q=1}=1$. This success comes from an ultimate identification process over
monomials like $(c_0)^p$. For osp(1$|$2), analogous principles of calculation are transposed,
involving a second parameter $d_0$. Ultimate identification process then must be done over 
binomials like ${(c_{0}+{d_{0}}^{2})}^{\Omega -m} \left({d_{0}}^{2}\right)^{m}$. {\em Unknown} 
polynomials ${\cal P}$ are introduced as well as their expansion coefficients, $x$, over the 
binomials. It is clearly shown that a hypothetical super-triangle sum rule requires super-triangles
$\bigtriangleup^{S}$, instead of $\bigtriangledown$ for su(2)/su$_q$(2). Coefficients $x$  are 
integers ({\em conjecture 1}). Massive unknown advances are done for intermediate steps of 
calculation. Among other, are proved {\em two  theorems} on tensor operators, ``zero" by 
construction. However, the ultimate identification seems to lead to a dead end, due to analytical
apparent complexities. Up today, except for a few of coefficients $x$, no general formula is really
available.

\end{abstract}

\newpage

\begin{flushleft}

                                                \section{Introduction}\label{Intro}
\renewcommand{\theequation}{\ref{Intro}.\arabic{equation}}
\setcounter{equation}{0}
\hspace*{1.5em}The well known ``triangle sum rule", or $\bigtriangledown$-sum rule, was first evidenced 
in 1971\cite{BiedLouck}. 
In our intrinsic approach, this identity between generic $6$-$j$ symbols results simply from an application of  so-called ``coupling laws" to special tensor operators, $\mbox{\boldmath$S$}^{\kappa}$, iterated  from a basic one with a rank $\frac{1}{2}$. Our method never need the heavy use of $3$-$j$ symbols,  but only $6$-$j$ symbols with one spin $\frac{1}{2}$. 
Efficiency of this point of view is easily ascertained for su(2) and su$_q$(2), but for osp(1$|$2), the method lets crop up unexpected difficulties regarding {\tt non-standard polynomial calculations}, analyzed in the main part of  this paper.\par
It is organized in the following way:\par
\hspace*{1.5em}{\em Section 2} introduces, for su(2), iterated tensor operators $\mbox{\boldmath$S$}^{\kappa}(c_0)$ depending on a 
real parameter $c_0$, built from a {\em fundamental} one, namely $\mbox{\boldmath$S$}^{\frac{1}{2}}(c_0)$ of rank
$\frac{1}{2}$. A closure relation for tensor products like 
$[\mbox{\boldmath$S$}^{a}\times \mbox{\boldmath$S$}^{b}]^{c}\propto\mbox{\boldmath$S$}^{c}$ is established in terms of  
$\bigtriangledown$-triangle, $\bigtriangledown(a b c)$.\par
\hspace*{1.5em}The short {\em section 3} uses customary su(2)-coupling laws for quickly deriving the well known triangle sum rule for su(2).\par
\hspace*{1.5em}{\em Section 4}, devoted to su$_q$(2), follows a similar way for obtaining a (new) compact expression for the 
$q$-triangle sum rule.\par
\hspace*{1.5em}{\em Section 5} consists in reminders related to the definitions of an osp(1$|$2)-supertriangle $\bigtriangleup^{S}$
and specific coupling laws for osp(1$|$2) tensor products.\par
\hspace*{1.5em}In {\em Section 6}, we show that, instead of only one parameter like for su(2) or su$_q$(2), two independent real variables $c_0$ and $d_0$ are necessary for defining iterated tensor operators $\mbox{\boldmath$S$}^{\kappa}(c_0,d_0)$. 
Analytical calculation of two relevant coefficients, $\alpha_{\kappa}(d_0)$ and $\gamma_{\kappa}(c_0+{d_ 0}^{2},{d_ 0}^{2})$, 
occuring in closure relations for osp(1$|$2) tensor products, is successfully carried out.\par
\hspace*{1.5em}{\em Section 7} contains an overview of the calculation method followed for obtaining at least {\em seven}
explicit closure relations for tensor products of highest ranks. {\tt \em {\small Theorem 1}} is proved, regarding a ``zero tensor product". \par 
\hspace*{1.5em}{\em Section 8} lets appear a first general closure relation for tensor operators, naturally, in terms of 
{\tt unavoidable} polynomials ${\cal{P}}^{\pi}(\lambda,\kappa)$, properly defined, and depending on $d_ 0,c_0+{d_ 0}^{2}$.
Actually these polynomials constitute the major trouble inherent in our theoretical approach because of the weak hope of finding a general formula for expansion coefficients $x(\lambda,\kappa)$ over the powers of $d_ 0$, $c_0+{d_ 0}^{2}$.
Coefficients $x$ are integers ({\tt \em {\small conjecture 1}}).\par
\hspace*{1.5em}{\em Sections 9-10} lead to similar results for tensor products of lowest ranks, depending in addition on the coefficients $\gamma_{\kappa}$. Instead of ${\cal{P}}^{\pi}(\lambda,\kappa)$, other ${\cal{Q}}^{\pi}(\lambda,\kappa)$ polynomials
are introduced, with ${\cal{Q}}^{\pi}(\lambda,\kappa)={\cal{P}}^{\pi}(\lambda,\kappa-\lambda+\pi)$ ({\tt \em {\small conjecture 2}}).
A {\tt \em {\small Theorem 2}} is proved, regarding another ``zero product".\par 
\hspace*{1.5em}{\em Section 11} outlines a single unified form of closure relation valid $\forall$ rank.\par
\hspace*{1.5em}In {\em Section 12}, one reproduces exactly the same steps of calculation as those successfully used for
su(2) or su$_q$(2). That matches the penultimate stage using triple tensor products before the final identification over
the specific osp(1$|$2)-parameters $d_ 0$ and $c_0+{d_ 0}^{2}$. The resulting relevant equations already confront us with an increasing degree of algebraic complexity in comparison with the analogous situation encountered for su(2).\par
\hspace*{1.5em}{\em Section 13}, a very important one, sets precisely all the polynomial definitions and conditions, which seem necessary for achieving the ultimate identification process over a lot of  binomials. However, in addition with a paradoxical remark about osp(1$|$2) (may be false), we give up finding an analytical solution to our problem, thus declared ``open". \par
\hspace*{1.5em}After the conclusion of the present work, three mathematical appendices are joined to the paper.\par
\hspace*{1.5em}{\em Appendix A} shows that reduced matrix elements of osp(1$|$2) tensor operators, like
$\mbox{\boldmath$S$}^{\frac{1}{2}}$ and $\mbox{\boldmath$S$}^{1}$, can be perfectly expressed by means of the coefficients   
$\alpha_{\kappa}$ and $\gamma_{\kappa}$.\par
\hspace*{1.5em}{\em Appendix B} contains a very detailed review of recursion relations regarding 
${\cal{P}}^{\omega}$ polynomials and their expansion coefficients ($x$) over binomials like
$\left(c_{0}+{d_{0}}^{2}\right)^{[\omega]-m}({d_{0}}^{2})^{m}$. Even, a few exact analytical expressions of some expansion coefficients have been listed. \par 

\hspace*{1.5em} In terms of parity-independent $6$-$j^{S}$ symbols, {\em Appendix C} re-actualizes a 20-old-year proposition
expounded for a peculiar osp(1$|$2)-triangle sum rule.  
 
    \section{Definition of iterated tensor operators $\mbox{\boldmath$S$}^{\kappa}$, from $\mbox{\boldmath$S$}^{\frac{1}{2}}$, for su(2)}\label{BasicDefs}
\renewcommand{\theequation}{\ref{BasicDefs}.\arabic{equation}}
\setcounter{equation}{0}For convenience, let us remind here the coupling laws for su(2) tensor operators, which will  be indispensable for our approach:\par
\hspace*{1.5em} {\em Left-recoupling:}
\begin{eqnarray}   \label{eq:SU2COUPLLAWlbis}
\lefteqn{ [\mbox{\boldmath$X$}^{j_{1}}\times
[\mbox{\boldmath$Y$}^{j_{2}}\times\mbox{\boldmath$Z$}^{j_{3}}]^{j_{23}}]^{j_{123}}=
(-1)^{j_{1}+j_2+j_3+j_{123}} (2j_{23}+1)^{\frac{1}{2}}  } \nonumber \\  
&& \times \sum_{j_{12}} (2j_{12}+1)^{\frac{1}{2}}
\left\{ \begin{array}{ccc} j_2 &j_3 & j_{23} \\
j_{123} & j_1 & j_{12} \end{array} \right\}
[[\mbox{\boldmath$X$}^{j_{1}}\times
\mbox{\boldmath$Y$}^{j_{2}}]^{j_{12}}\times\mbox{\boldmath$Z$}^{j_{3}}]^{j_{123}}. 
\end{eqnarray}
\hspace*{1.5em} {\em Right-recoupling:}
\begin{eqnarray}   \label{eq:SU2COUPLLAWrbis}
\lefteqn{ [[\mbox{\boldmath$X$}^{j_{1}}\times
\mbox{\boldmath$Y$}^{j_{2}}]^{j_{12}}\times\mbox{\boldmath$Z$}^{j_{3}}]^{j_{123}}= 
(-1)^{j_1+j_2+j_3+j_{123}}(2j_{12}+1)^{\frac{1}{2}} } \nonumber \\ 
&& \times  \sum_{j_{23}} (2j_{23}+1)^\frac{1}{2}
\left\{ \begin{array}{ccc} j_1 &j_2 & j_{12} \\
j_3 & j_{123} & j_{23} \end{array} \right\}
[\mbox{\boldmath$X$}^{j_{1}}\times
[\mbox{\boldmath$Y$}^{j_{2}}\times\mbox{\boldmath$Z$}^{j_{3}}]^{j_{23}}]^{j_{123}}. 
\end{eqnarray}
Also it's assumed the knowledge of tables for $6$-$j$ symbols with one argument equal to $0$ or $\frac{1}{2}$ \cite{Edmonds}, the only ones to be used in our analysis. \par
Now we can proceed to the definition of $\mbox{\boldmath$S$}^{\kappa}$.\par

{\tt Only two defining equations are necessary, namely}
\begin{equation}   \label{eq:STensOpwithrank0andk}
[\mbox{\boldmath$S$}^{\frac{1}{2}}\times\mbox{\boldmath$S$}^{\frac{1}{2}}]^{0}=c_{0},
\mbox{\hspace{0.5em}and\hspace{0.5em}}[\mbox{\boldmath$S$}^{\frac{1}{2}}\times\mbox{\boldmath$S$}^{\kappa}]^{\kappa+\frac{1}{2}}=\mbox{\boldmath$S$}^{\kappa+\frac{1}{2}}. 
\end{equation}
In the present work, $c_{0}$ will be assumed to be a real number, {\em i.e.} neither a pure imaginary nor a complex.\par 
Thanks to recoupling equations and tables aforementioned, it follows that:
\begin{equation}   \label{eq:STenswithrank0andSymkhalf}
[\mbox{\boldmath$S$}^{\frac{1}{2}}\times\mbox{\boldmath$S$}^{\kappa}]^{\kappa+\frac{1}{2}}=
[\mbox{\boldmath$S$}^{\kappa}\times\mbox{\boldmath$S$}^{\frac{1}{2}}]^{\kappa+\frac{1}{2}}=
\mbox{\boldmath$S$}^{\kappa+\frac{1}{2}}
\mbox{\hspace{0.5em}and\hspace{0.5em}} \mbox{\boldmath$S$}^{0}=\mbox{\boldmath$1$}^{0}. 
\end{equation}
A reasoning by recursion allows one that   the following property holds:
\begin{equation}   \label{eq:Coeffgammak}
[\mbox{\boldmath$S$}^{\frac{1}{2}}\times\mbox{\boldmath$S$}^{\kappa}]^{\kappa-\frac{1}{2}}=
[\mbox{\boldmath$S$}^{\kappa}\times\mbox{\boldmath$S$}^{\frac{1}{2}}]^{\kappa-\frac{1}{2}}=
\gamma_{\kappa} \mbox{\boldmath$S$}^{\kappa-\frac{1}{2}}.
\end{equation}

                              \subsection{Detailed calculation of $\gamma_{\kappa}$}\label{gammak} 
Consider first the following equation:
\begin{equation} \label{eq:c0S1/2}
[\mbox{\boldmath$S$}^{\frac{1}{2}}\times[\mbox{\boldmath$S$}^{\frac{1}{2}}\times\mbox{\boldmath$S$}^{\frac{1}{2}}]^{0}]^{\frac{1}{2}}=
[\mbox{\boldmath$S$}^{\frac{1}{2}}\times c_{0}\mbox{\boldmath$1$}^{0}]^{\frac{1}{2}}=         
c_{0}\mbox{\boldmath$S$}^{\frac{1}{2}}.
\end{equation}
Left-recoupling and use of the value of $6$-$j$ symbols with one argument $0$ lead to the value of $\gamma_{1}$. The same result holds
if one works with eq.~(\ref{eq:c0S1/2}) easily written in an alternative way and a right-recoupling. That yields:
\begin{equation}   \label{eq:gamma1}
[\mbox{\boldmath$S$}^{\frac{1}{2}}\times \mbox{\boldmath$S$}^{1}]^{\frac{1}{2}}=
[ \mbox{\boldmath$S$}^{1}\times\mbox{\boldmath$S$}^{\frac{1}{2}}]^{\frac{1}{2}}=
c_{0}\sqrt{3}\mbox{\boldmath$S$}^{\frac{1}{2}}=\gamma_{1}\mbox{\boldmath$S$}^{\frac{1}{2}}.
\end{equation}
The same method using a reasoning by recursion according to eq.~(\ref{eq:Coeffgammak}) may be applied in considering the product 
$[\mbox{\boldmath$S$}^{\frac{1}{2}}\times \mbox{\boldmath$S$}^{\kappa+\frac{1}{2}}]^{\kappa}=\gamma_{\kappa+\frac{1}{2}}\mbox{\boldmath$S$}^{\kappa+\frac{1}{2}}=
[\mbox{\boldmath$S$}^{\frac{1}{2}}\times[\mbox{\boldmath$S$}^{\kappa}\times\mbox{\boldmath$S$}^{\frac{1}{2}}]^{\kappa+\frac{1}{2}}]^{\kappa}$ and a left-recoupling.
Use of expressions of the $6$-$j$ symbols with one argument $\frac{1}{2}$ leads finally to the general formula of $\gamma_{\kappa}$:
\begin{equation}   \label{eq:Finalgammak}
\gamma_{\kappa}=c_{0}\left[ \frac{(2\kappa+1)2\kappa}{2}\right]^{\frac{1}{2}}.
\end{equation}
We have now at our's disposal complete analytical expressions for tensor products involving one $\mbox{\boldmath$S$}^{\frac{1}{2}}$ at least.
This is sufficient for deducing by a recursive analysis that the following important property holds: 
\begin{equation}   \label{eq:symmetrySkSkprime}
[\mbox{\boldmath$S$}^{\kappa}\times \mbox{\boldmath$S$}^{\kappa'}]^{\kappa''}=
[\mbox{\boldmath$S$}^{\kappa'}\times \mbox{\boldmath$S$}^{\kappa}]^{\kappa''}. 
\end{equation}

             \subsection{General formula for $[\mbox{\boldmath$S$}^{\lambda}\times \mbox{\boldmath$S$}^{\kappa}]^{\lambda +\kappa-1}$}\label{Formulaforp}
Once easily checked that
\begin{equation}   \label{eq:Casep0}
[\mbox{\boldmath$S$}^{\lambda}\times \mbox{\boldmath$S$}^{\kappa}]^{\lambda+\kappa}=
\mbox{\boldmath$S$}^{\lambda+\kappa},
\end{equation}
one starts for instance with the following expression:
\begin{equation}   \label{eq:Generallamdapluskminus1}
[\mbox{\boldmath$S$}^{\lambda}\times \mbox{\boldmath$S$}^{\kappa}]^{\lambda +\kappa-1}=
[[\mbox{\boldmath$S$}^{\lambda-\frac{1}{2}}\times \mbox{\boldmath$S$}^{\frac{1}{2}}]^{\lambda}\times\mbox{\boldmath$S$}^{\kappa}]^{\lambda +\kappa-1}.
\end{equation}
Left-recoupling, use of values for $6$-$j$ symbols involved and eq.~(\ref{eq:Finalgammak}) lead to
\begin{equation}   \label{eq:SlamdaSkSlamdaplusk-1}
[\mbox{\boldmath$S$}^{\lambda}\times \mbox{\boldmath$S$}^{\kappa}]^{\lambda +\kappa-1}=
c_{0}\left[ \frac{2\lambda\cdot 2\kappa(2\kappa+2\lambda)}{2}\right]^{\frac{1}{2}}
\mbox{\boldmath$S$}^{\lambda +\kappa-1}.
\end{equation}
 
        \subsection{Recursion relation for $[\mbox{\boldmath$S$}^{\lambda}\times \mbox{\boldmath$S$}^{\kappa}]^{\lambda +\kappa-p}$}\label{Recurslambdap}
From the preceding results, for example a right-recoupling on
$[\mbox{\boldmath$S$}^{\lambda}\times \mbox{\boldmath$S$}^{\kappa}]^{\lambda +\kappa-p}=
[[\mbox{\boldmath$S$}^{\lambda-\frac{1}{2}}\times\mbox{\boldmath$S$}^{\frac{1}{2}}]^{\lambda}\times 
\mbox{\boldmath$S$}^{\kappa}]^{\lambda +\kappa-p}$ yields:
\begin{eqnarray}   \label{eq:RecursRelforp}
\left[\scriptstyle\frac{(2\lambda)!(2\kappa-p)!}{(2\lambda -p)!(2\kappa)!}\right]^{\frac{1}{2}}
[\mbox{\boldmath$S$}^{\lambda}\times \mbox{\boldmath$S$}^{\kappa}]^{\lambda +\kappa-p}= \mbox{\hspace{17em}}
\nonumber\\
\left( \frac{c_0}{\sqrt{2}}\right) 
\left[\scriptstyle\frac{p(2\lambda+2\kappa-p+1)(2\lambda -1)!(2\kappa-p)!}{(2\lambda -p)!(2\kappa-1)!}\right]^{\frac{1}{2}}
[\mbox{\boldmath$S$}^{\lambda-\frac{1}{2}}\times 
\mbox{\boldmath$S$}^{\kappa-\frac{1}{2}}]^{(\lambda-\frac{1}{2})+(\kappa-\frac{1}{2})-(p-1)}  \nonumber \\
+ \left[\scriptstyle\frac{(2\lambda -1)!(2\kappa-p+1)!}{(2\lambda -p-1)!(2\kappa+1)!}\right]^{\frac{1}{2}}
[\mbox{\boldmath$S$}^{\lambda-\frac{1}{2}}\times \mbox{\boldmath$S$}^{\kappa+\frac{1}{2}}]^{\lambda +\kappa-p}.\nonumber \\
\end{eqnarray}
By introducing a coefficient $\beta$ (to be determined) with the following equation:		
\begin{equation}   \label{eq:CoeffBetaIntro}
\frac{1}{\sqrt{p!}}
\left( \frac{\sqrt{2}}{c_0}\right)^{p}
\left[\scriptstyle \frac{(2\kappa+2\lambda -2p+1)!}{(2\kappa+2\lambda -p+1)!}\right]^{\frac{1}{2}}
\left[{\scriptstyle\frac{(2\lambda)!(2\kappa-p)!}{(2\lambda -p)!(2\kappa)!}}\right]^{\frac{1}{2}}
[\mbox{\boldmath$S$}^{\lambda}\times \mbox{\boldmath$S$}^{\kappa}]^{\lambda +\kappa-p}=
\beta_{p}^{2\lambda,\kappa}\mbox{\boldmath$S$}^{\lambda +\kappa-p},
\end{equation}
one deduces that
\begin{equation}   \label{eq:RecursBeta}
\beta_{p}^{2\lambda,\kappa}=\beta_{p-1}^{2\lambda -1,\kappa}+\beta_{p}^{2\lambda -1,\kappa}.
\end{equation}
Eq.~(\ref{eq:SlamdaSkSlamdaplusk-1}) gives $\beta_{1}^{2\lambda,k}=2\lambda$. Then, after noting the symmetry $\lambda\leftrightarrow \kappa$, it can be seen that coefficient $\beta$ does not depend on $\kappa$. Eq.~(\ref{eq:RecursBeta}) becomes simply a well known recursion relation for binomial coefficients, whence
\begin{equation}   \label{eq:ExpressBeta}
\beta_{p}^{2\lambda}=\frac{(2\lambda)!}{(2\lambda -p)!p!}.
\end{equation}
The final result for $[\mbox{\boldmath$S$}^{\lambda}\times \mbox{\boldmath$S$}^{\kappa}]^{\lambda +\kappa-p}$  then reads:
\begin{equation}   \label{eq:FinalResSlamdakp}
[\mbox{\boldmath$S$}^{\lambda}\times \mbox{\boldmath$S$}^{\kappa}]^{\lambda +\kappa-p}=
\left( \frac{c_0}{\sqrt{2}}\right)^{p}
\left[\scriptstyle\frac{(2\kappa)!(2\lambda)!(2\kappa+2\lambda -p+1)!}
{p!(2\kappa-p)!(2\lambda-p)!(2\kappa+2\lambda -2p+1)!}\right]^{\frac{1}{2}}\mbox{\boldmath$S$}^{\lambda +\kappa-p}.
\end{equation}
After a variables change $\lambda \rightarrow a, \kappa \rightarrow b, \lambda +\kappa-p \rightarrow c$, 
the preceding equation can be rewritten as a {\em closure relation}:
\begin{equation}   \label{eq:ClosureRelationsu2}
[\mbox{\boldmath$S$}^{a}\times \mbox{\boldmath$S$}^{b}]^{c}=
\left(\frac{c_0}{\sqrt{2}}\right)^{a+b-c}
\left[\frac{(2a)!(2b)!}{(2c+1)!}\right]^{\frac{1}{2}} \bigtriangledown(a b c) \mbox{\boldmath$S$}^{c},
\end{equation}
where the ``triangle coefficient" $\bigtriangledown$ is given by the inverse of the well known $\bigtriangleup$ triangle
\cite{Edmonds}, namely
\begin{equation}   \label{eq:TriangEdmdown}
\bigtriangledown(a b c)=
\left[ \frac{(a+b+c+1)!}{(a+b-c)!(a-b+c)!(-a+b+c)!} \right]^{1/2}.
\end{equation} 
 
                                                 \section{Triangle sum rule for su(2)}\label{Finalsu2}
\renewcommand{\theequation}{\ref{Finalsu2}.\arabic{equation}}
\setcounter{equation}{0}
Consider the triple product $[[\mbox{\boldmath$S$}^{a}\times \mbox{\boldmath$S$}^{b}]^{c}\times\mbox{\boldmath$S$}^{d}]^{e}$.

                                                 \subsection{A first expression of the triple product}
From the closure relation (\ref{eq:ClosureRelationsu2}), it can be obtained a first expression:
\begin{eqnarray}   \label{eq:SaSbrankcSdrankdranke}
\lefteqn{
[ [\mbox{\boldmath$S$}^{a}\times \mbox{\boldmath$S$}^{b}]^{c}\times\mbox{\boldmath$S$}^{d} ]^{e}=
\left(\frac{c_0}{\sqrt{2}}\right)^{a+b+d-e} } \nonumber \\
&& \times \left[\frac{(2a)!(2b)!}{(2c+1)!}\cdot\frac{(2c)!(2d)!}{(2e+1)!}\right]^{\frac{1}{2}}
\bigtriangledown(a b c)\bigtriangledown(c d e) \mbox{\boldmath$S$}^{e}.
\end{eqnarray}
 
                                       \subsection{A second expression from tensorial recoupling law}\label{SecondExpress}
One carries out a right-recoupling over the triple product according to eq.~(\ref{eq:SU2COUPLLAWrbis}), then one uses again
the closure property (\ref{eq:ClosureRelationsu2}), that gives:
\begin{eqnarray}   \label{eq:SaSbrankcSdrankdrankebis}
\lefteqn{
[ [\mbox{\boldmath$S$}^{a}\times \mbox{\boldmath$S$}^{b}]^{c}\times\mbox{\boldmath$S$}^{d} ]^{e}=
(-1)^{a+b+d+e} \left(\frac{c_0}{\sqrt{2}}\right)^{a+b+d-e} } \nonumber \\
&& \times \sqrt{(2c+1)} \left[\frac{(2b)!(2d)!(2a)!}{(2e+1)!}\right]^{\frac{1}{2}}
\displaystyle \sum_{f}  
\left\{ \begin{array}{ccc} a & b & c \\d & e & f \end{array} \right\}
\bigtriangledown(b d f)\bigtriangledown(a f e) \mbox{\boldmath$S$}^{e}.
\end{eqnarray}
   
                                         \subsection{Identification of both expressions}\label{Identifboth}
Identification of eq.~(\ref{eq:SaSbrankcSdrankdranke}) and eq.~(\ref{eq:SaSbrankcSdrankdrankebis}) over $\mbox{\boldmath$S$}^{e}$ furnishes
the expected result:
\begin{equation}   \label{eq:TrianglSumRulesu(2)}
\bigtriangledown(a b c)\bigtriangledown(c d e)=(-1)^{a+b+d+e}
(2c+1)\displaystyle \sum_{f}
\bigtriangledown(b d f)\bigtriangledown(a f e)  
\left\{ \begin{array}{ccc} a & b & c \\d & e & f \end{array} \right\}.
\end{equation}
This is exactly the triangle sum rule for su(2).\par
\underline{{\em Remark:}}\par
\hspace*{1.5em}The standard generator for su(2), {\em i.e.} the angular momentum itself $\mbox{\boldmath$J$}^{1}$, satisfies the following commutation relations: 
\begin{equation}   \label{eq:Commutsu2}
[\mbox{\boldmath$J$}^{1}\times\mbox{\boldmath$J$}^{1}]^{1}
=-\frac{1}{\sqrt{2}}\mbox{\boldmath$J$}^{1}.
\end{equation}
Looking at eq.~(\ref{eq:SlamdaSkSlamdaplusk-1}), it can be seen that {\em only one} tensor operator of rank $1$, namely
$\mbox{\boldmath$S$}^{1}$, actually depending on $c_{0}$, satisfies commutation relations such as given by eq.~(\ref{eq:Commutsu2})
if we set $c_{0}=-\frac{1}{4}$. Therefore ${\mbox{\boldmath$S$}^{1}}_{|c_{0}=-\frac{1}{4}}$ is another generator for su(2), in this instance, according to the terminology of ref.~\cite{BiedLouck}, the ``symplectic" generator for su(2).

                                          \section{$q$-Triangle sum rule for su$_q$(2)}\label{Finalsuq2}
\renewcommand{\theequation}{\ref{Finalsuq2}.\arabic{equation}}
\setcounter{equation}{0}

\hspace*{1.5em} In a recent paper \cite[pp. 375-376]{L.B.Nuovo.III}, we have properly defined $q$-irreducible tensor operators and their tensorial products.
From that, it can be deduced that $q$-coupling laws for $q$-irreducible tensor operators are formally identical to those of su(2) (apart from the occurrence of $q$ and the change of usual numbers into $q$-numbers)). The formulas are written down below.\par
\hspace*{1.5em} {\em $q$-Left-recoupling:}
\begin{eqnarray}   \label{eq:qSU2COUPLLAWlbis}
\lefteqn{ [\mbox{\boldmath$X$}^{j_{1}}\times
[\mbox{\boldmath$Y$}^{j_{2}}\times\mbox{\boldmath$Z$}^{j_{3}}]^{j_{23}}]^{j_{123}}=
(-1)^{j_{1}+j_2+j_3+j_{123}} \sqrt{[2j_{23}+1]}  } \\ 
&& \times \sum_{j_{12}} \sqrt{[2j_{12}+1]}
\left\{ \begin{array}{ccc} j_2 &j_3 & j_{23} \\
j_{123} & j_1 & j_{12} \end{array} \right\}_{q}
[[\mbox{\boldmath$X$}^{j_{1}}\times
\mbox{\boldmath$Y$}^{j_{2}}]^{j_{12}}\times\mbox{\boldmath$Z$}^{j_{3}}]^{j_{123}}. \nonumber
\end{eqnarray}
\hspace*{1.5em} {\em $q$-Right-recoupling:}
\begin{eqnarray}   \label{eq:qSU2COUPLLAWrbis}
\lefteqn{ [[\mbox{\boldmath$X$}^{j_{1}}\times
\mbox{\boldmath$Y$}^{j_{2}}]^{j_{12}}\times\mbox{\boldmath$Z$}^{j_{3}}]^{j_{123}}= 
(-1)^{j_1+j_2+j_3+j_{123}}\sqrt{[2j_{12}+1]} } \nonumber \\ 
&& \times  \sum_{j_{23}} \sqrt{[2j_{23}+1]}
\left\{ \begin{array}{ccc} j_1 &j_2 & j_{12} \\
j_3 & j_{123} & j_{23} \end{array} \right\}_{q}
[\mbox{\boldmath$X$}^{j_{1}}\times
[\mbox{\boldmath$Y$}^{j_{2}}\times\mbox{\boldmath$Z$}^{j_{3}}]^{j_{23}}]^{j_{123}}. 
\end{eqnarray}
Definitions/property of $q$-iterated tensor operators $\mbox{\boldmath$S$}^{\kappa}$ are exactly the same as those expressed by 
eqs.~(\ref{eq:STensOpwithrank0andk})-(\ref{eq:Coeffgammak}). On the other hand eq.~(\ref{eq:c0S1/2}) transforms into   
\begin{equation}   \label{eq:gammaq1}
[\mbox{\boldmath$S$}^{\frac{1}{2}}\times \mbox{\boldmath$S$}^{1}]^{\frac{1}{2}}=
[ \mbox{\boldmath$S$}^{1}\times\mbox{\boldmath$S$}^{\frac{1}{2}}]^{\frac{1}{2}}=
c_{0}\frac{(1+[2])}{\sqrt{[3]}}
\mbox{\boldmath$S$}^{\frac{1}{2}}
\equiv c_{0}\frac{\sqrt{[3]}}{([2]-1)}\mbox{\boldmath$S$}^{\frac{1}{2}},
\end{equation}
whence, for su$_{q}$(2), 
\begin{equation}   \label{eq:gammaq1final}
\gamma_{1}=c_{0}\frac{\sqrt{[3]}}{([2]-1)}.
\end{equation}
Recursion relation for $\gamma_{\kappa}$ is found to be
\begin{equation}   \label{eq:Recursgammaqk}
\gamma_{\kappa+\frac{1}{2}}([2\kappa+1]-1)=\sqrt{[2\kappa+2][2\kappa]} \gamma_{\kappa},
\end{equation}
with the following solution:
\begin{equation}   \label{eq:formulagammaqk}
\gamma_{\kappa}=
\sqrt{\frac{[2\kappa+1][2\kappa]}{[2]}}
\frac{[2\kappa-1]!}{([2\kappa]-1)\cdots([2]-1)}c_0. 
\end{equation}
Coefficients $\gamma_{\kappa}$ can be expressed as a function of the series $F(n)$ studied in the paper of Nomura and Biedenharn \cite{NomurBied}.
\begin{equation}   \label{eq:Fnseries}
F(n)=[1]+[2]+\cdots +[n],
\end{equation}
having the following recursive property:
\begin{equation}   \label{eq:recursFn}
F(n)=\frac{[n+1]}{[n]-1} F(n-1).
\end{equation}
Thus $\gamma_{\kappa}$ may be rewritten under the form
\begin{equation}   \label{eq:gammakqFn}
\gamma_{\kappa}=
\sqrt{\frac{[2\kappa+1][2\kappa]}{[2]}} \frac{[2]F(2\kappa)}{[2\kappa+1][2\kappa]} c_0.
\end{equation} 
\hspace*{1.5em}However, for obvious reasons  regarding easy comparisons with the su(2) case with $q=1$, we shall use the following definition of a
series $\Phi(n)$:
\begin{equation}   \label{eq:DefofPhi}
\Phi(n)=\frac{[2]F(2\kappa)}{[2\kappa+1][2\kappa]} \mbox{\hspace{4em}(defining equation of series $\Phi$}).
\end{equation}
Clearly we have
\begin{equation}   \label{eq:Philimitq1}
\Phi(n)_{| q=1}=1.
\end{equation}
Thus $\gamma_{\kappa}$ has the form:
\begin{equation}   \label{eq:gammaqkPhi}
\gamma_{\kappa} = c_0\sqrt{\frac{[2\kappa+1][2\kappa]}{[2]}} \Phi{(2\kappa)} .
\end{equation}
Analogy with eq. (\ref{eq:Finalgammak}) thus is immediate.\par
Like in ref. \cite{NomurBied}, we can define a factorial of $\Phi$ by
\begin{equation}   \label{eq:PhiFactorial}
\Phi(n)!=\Phi(n)\Phi(n-1)\cdots\Phi(1).
\end{equation}
By using eq.~(\ref{eq:Generallamdapluskminus1}) like for su(2), {\em and noting that this chosen form implies that $\lambda=\inf(\lambda,\kappa)$ - a fact of importance allowing the iterations to come -}, one sees that the recoupling process leads to a basic recursion relation:
\begin{eqnarray}   \label{eq:qRecursRelforp}
[\mbox{\boldmath$S$}^{\lambda}\times \mbox{\boldmath$S$}^{\kappa}]^{\lambda +\kappa-p}= \mbox{\hspace{17em}} \nonumber \\
\sqrt{\frac{[2\kappa+2\lambda-p+1][p]}{[2\kappa+1][2\lambda]}} \gamma_{\kappa}
[\mbox{\boldmath$S$}^{\lambda-\frac{1}{2}}\times \mbox{\boldmath$S$}^{\kappa-\frac{1}{2}}]^{(\lambda-\frac{1}{2})+(\kappa-\frac{1}{2})-(p-1)}  \nonumber \\
+ \sqrt{\frac{[2\lambda-p][2\kappa-p+1]}{[2\kappa+1][2\lambda]}}
[\mbox{\boldmath$S$}^{\lambda-\frac{1}{2}}\times \mbox{\boldmath$S$}^{\kappa+\frac{1}{2}}]^{\lambda +\kappa-p}.
\end{eqnarray}
If $p=1$, thanks to eq.~(\ref{eq:Casep0}), this latter equation is easily solved by iterating with $\lambda \rightarrow \lambda-\frac{1}{2}$, 
$\kappa \rightarrow \kappa+\frac{1}{2}$ and so on. One finds:
\begin{equation}   \label{eq:rankwithp1}
\sqrt{\frac{[2\lambda]}{[2\kappa]}}
[\mbox{\boldmath$S$}^{\lambda}\times \mbox{\boldmath$S$}^{\kappa}]^{\lambda +\kappa-1}=
\sqrt{\frac{[2\kappa+2\lambda]}{[2]}}
\left(\sum_{n=0}^{n=2\lambda-1} \Phi(2\kappa+n)\right) \mbox{\boldmath$S$}^{\lambda+\kappa-1}.
\end{equation}
Note that a fully symmetrical formula in $\lambda, \kappa$ can be written simply by the following replacement
$\lambda \rightarrow \inf(\lambda,\kappa)$, $\kappa \rightarrow \sup(\lambda,\kappa)$.\par
\hspace*{1.5em}Let us adopt the simplest expression possible for the $q$-analog of the triangle coefficient, {\em i.e.}:
\begin{equation}   \label{eq:qTriangledown}
\bigtriangledown_q(a b c)=
\left[ \frac{[a+b+c+1]!}{[a+b-c]![a-b+c]![-a+b+c]!} \right]^{1/2}.
\end{equation}
Analogously to the su(2) closure relation~(\ref{eq:ClosureRelationsu2}), {\tt we define a coefficient $\omega$}, {\tt symmetrical in $(a,b)$}
by means of  the following equation:
\begin{equation}   \label{eq:qClosureRelationsu2}
[\mbox{\boldmath$S$}^{a}\times \mbox{\boldmath$S$}^{b}]^{c}=\omega_{a+b-c}^{a,b}
\left(\frac{c_0}{\sqrt{[2]}}\right)^{a+b-c}
\left[\frac{[2a]![2b]!}{[2c+1]!}\right]^{\frac{1}{2}} \bigtriangledown_q(a b c) \mbox{\boldmath$S$}^{c}.
\end{equation}
According to eq.~(\ref{eq:Casep0}), obviously valid for su$_q$(2), we have:
\begin{equation}   \label{eq:omegap0}
\omega_{0}^{a,b}=1.
\end{equation}
It remains now to determine a precise analytical expression of $\omega$. This can be done from the recursion relation~(\ref{eq:qRecursRelforp}).
As a function of $\omega$, it becomes:
\begin{equation}    \label{eq:qRecursforomega}
{ \scriptstyle
\left[ \begin{array}{c} 2\lambda \\ p \end{array} \right] }
\omega_{p}^{\lambda,\kappa}=
\Phi(2\kappa) { \scriptstyle
\left[ \begin{array}{c} 2\lambda-1 \\ p-1 \end{array} \right] } \omega_{p-1}^{\lambda-\frac{1}{2},\kappa-\frac{1}{2}}
+ { \scriptstyle \left[ \begin{array}{c} 2\lambda-1 \\ p \end{array} \right] } \omega_{p}^{\lambda-\frac{1}{2},\kappa+\frac{1}{2}}.
\end{equation}
Note that this equation can be viewed as a poised formula of the su$_q$(2) identity between $q$-binomial coefficients, namely
\begin{equation}   \label{eq:Identityqbinomial}
\left[ \begin{array}{c} 2\lambda \\ p \end{array} \right]=\left[ \begin{array}{c} 2\lambda-1 \\ p-1 \end{array} \right]
+ \left[ \begin{array}{c} 2\lambda-1 \\ p \end{array} \right].
\end{equation}
The su(2) case with $q=1$ corresponds to:
\begin{equation}   \label{eq:omegasu2equal1}
{\omega_{p}^{\lambda,\kappa}}_{|q=1}=1.
\end{equation} 
A first interesting step is to examine eq.~(\ref{eq:qRecursforomega}) when $p=2\lambda$, because, in this case, 
${ \left[ \begin{array}{c} 2\lambda-1 \\ p \end{array} \right] }=0$.  \par \vspace{0.7em}
The result is:
\begin{equation}   \label{eq:omega2lambda}
\omega_{2\lambda}^{\lambda,\kappa}=\frac{\Phi(2\kappa)!}{\Phi(2\kappa-2\lambda)!}.
\end{equation}
The symmetrical form in $\lambda,\kappa$ is given by:
\begin{equation}   \label{eq:omega2lambdasym}
\omega_{2\inf(\lambda,\kappa)}^{\lambda,\kappa}=\frac{\Phi(2\sup(\lambda,\kappa)!}{\Phi(2|\kappa-\lambda|)!}.
\end{equation}
This shows that factorial definitions such as $\Phi(n)!$ (or $F(n)!$) are strongly 
related to $q$-scalar products like $[\mbox{\boldmath$S$}^{\lambda}\times \mbox{\boldmath$S$}^{\lambda}]^{0}$.

By iterating the recursion relation (\ref{eq:qRecursforomega}) with $\lambda \rightarrow \lambda-\frac{1}{2}$, 
$\kappa \rightarrow \kappa+\frac{1}{2}$, one finds:
\begin{equation}   \label{eq:qRecursRelomega}
{ \scriptstyle
\left[ \begin{array}{c} 2\lambda \\ p \end{array} \right] }
\omega_{p}^{\lambda,\kappa}=
\sum_{m=0}^{m=2\lambda-p} \Phi(2\kappa+m) { \scriptstyle
\left[ \begin{array}{c} 2\lambda - (m+1)\\ p-1 \end{array} \right] } 
\omega_{p-1}^{\lambda-\frac{(m+1)}{2},\kappa+\frac{(m-1)}{2}}.
\end{equation}
In terms of  coefficients $\omega$, eq.~(\ref{eq:rankwithp1}) may be rewritten as follows:
\begin{equation}   \label{eq:omegalambdakappa1}
{ \scriptstyle
\left[ \begin{array}{c} 2\lambda \\ 1 \end{array} \right] }
\omega_{1}^{\lambda,\kappa}=\sum_{m=0}^{m=2\lambda-1} \Phi(2\kappa+m).
\end{equation}
\hspace*{1.5em}In contrast to the method followed by Nomura and Biedenharn for obtaining their impressive formula, see eqs. (7.17)-( 7.18), p. 3645 
in ref.~\cite{NomurBied}, we choose to carry out iterations on eq.~(\ref{eq:qRecursRelomega}) from the top of the ranks $\lambda+\kappa-p$, related to tensor products
$[\mbox{\boldmath$S$}^{\lambda}\times \mbox{\boldmath$S$}^{\kappa}]^{\lambda +\kappa-p}$, with $p=1,2, \cdots, 2\lambda$, instead of the bottom starting from
$p=2\lambda$.\par
Our result is specially simple:
\begin{equation}   \label{eq:qomegafinal}
{ \scriptstyle
\left[ \begin{array}{c} 2\lambda \\ p \end{array} \right] }
\omega_{p}^{\lambda,\kappa}=
 \prod _{l=0}^{p-1} \;
\sum _{m_{l+1}=m_{l}}^{2\lambda-p} \Phi\bigl(2\kappa+m_{l+1}-l\bigr),
\end{equation}
valid $\forall$ $p$ $\epsilon$  $[1,2\lambda]$,  and with the convention $m_0=0$.\par 
It can be noticed that, if $p=2\lambda$ then $m_l=0$ $ \forall$ $l$ (always in the case $\lambda = \inf(\lambda,\kappa)$), one retrieves eq.~(\ref{eq:omega2lambda}).\par
Fully symmetrical formula in $a,b$ for $\omega$ thus is given by:
\begin{equation}   \label{eq:qomegafinalsym}
{ \scriptstyle
\left[ \begin{array}{c} a+b-|a-b| \\ a+b-c \end{array} \right] }
\omega_{a+b-c}^{a,b}=
 \prod _{l=0}^{a+b-c-1} \;
\sum _{m_{l+1}=m_{l}}^{c-|a-b|} \Phi\bigl(a+b+|a-b|+m_{l+1}-l\bigr),
\end{equation}
valid $ \forall$ $ (a+b-c) \geq 1$, and with the convention $m_0=0$.\par 
An ultimate step is to ``transcript" our recoupling method described in sect.~\ref{Finalsu2}, then the final expression
 for the $\mbox{\boldmath$\bigtriangledown_q$}$-{\bf triangle sum rule} may be written as follows:
\begin{eqnarray}   \label{eq:TrianglSumRulesuq(2)}
\lefteqn {\omega_{a+b-c}^{a,b}\;\omega_{c+d-e}^{c,d}\bigtriangledown_{q}(a b c)\bigtriangledown_{q}(c d e) } \nonumber \\
&& = (-1)^{a+b+d+e}
[2c+1]\displaystyle \sum_{f}
\omega_{b+d-f}^{b,d}\;\omega_{a+f-e}^{a,f}\bigtriangledown_{q}(b d f)\bigtriangledown_{q}(a f e)  
\left\{ \begin{array}{ccc} a & b & c \\d & e & f \end{array} \right\}_{q}. \nonumber \\
\end{eqnarray}

                                          \section{Generalization to osp(1$|$2)}\label{osp12Begin}
\renewcommand{\theequation}{\ref{osp12Begin}.\arabic{equation}}
\setcounter{equation}{0}
  
Expression of coupling laws and table of primitive $6$-$j^S$ symbols with arguments $0$ or $\frac{1}{2}$ can be found in our recent paper
\cite{L.B.Nuovo.I}. For convenience we transcript below the formulas for tensor products which show the use of integral parts, denoted by
$[\cdots]$
                                      \footnote{Although of common use, this notation, unfortunately similar to the one used for $q$-numbers, should not lead to any confusions in the sequel of the present paper.}
.
   \subsection{Reminders of some properties of osp(1$|$2)}\label{Remindosp12}
First we give the osp(1$|$2) analog of $\bigtriangledown$, which is called $\bigtriangledown^{S}$ \cite{L.B.Nuovo.I}:
\begin{equation}   \label{eq:superdeltainv}
\bigtriangledown^{S}(a b c)=\left[ \frac{[a+b+c+\frac{1}{2}]!}{[a+b-c]![a-b+c]![-a+b+c]!}\right]^{\frac{1}{2}}.
\end{equation}
It is useful also to transcript our definition of the supertriangle $\bigtriangleup^{S}$ \cite{L.B.Nuovo.I}, because, unexpectedly,  it will be an essential
parameter of our present study:
\begin{equation}   \label{eq:superdeltatriangle}
\bigtriangleup^{S}(a b c)=\left[\frac{[a+b-c]![a-b+c]![-a+b+c]!}{[a+b+c+\frac{1}{2}]!}\right]^{\frac{1}{2}}.
\end{equation}
 
Recoupling laws for tensor products of tensor operators are the following:\par
\hspace*{1.5em} {\em Left-recoupling:}
\begin{eqnarray}   \label{eq:OSP12COUPLLAWSlbis}
[\mbox{\boldmath$X$}^{J_{1}}\times
[\mbox{\boldmath$Y$}^{J_{2}}\times\mbox{\boldmath$Z$}^{J_{3}}]^{J_{23}}]^{J_{123}}=
(-1)^{[J_2+J_3+J_{23}]+[J_{1}+J_{23}+J_{123}]+2J_{23}}   \nonumber \\ 
\times \sum_{J_{12}}
\left\{ \begin{array}{ccc} J_2 &J_3 & J_{23} \\
J_{123} & J_1 & J_{12} \end{array} \right\}^{\!_S}
[[\mbox{\boldmath$X$}^{J_{1}}\times
\mbox{\boldmath$Y$}^{J_{2}}]^{J_{12}}\times\mbox{\boldmath$Z$}^{J_{3}}]^{J_{123}}.
\end{eqnarray}
\hspace*{1.5em} {\em Right-recoupling:}
\begin{eqnarray}   \label{eq:OSP12COUPLLAWSrbis}
[[\mbox{\boldmath$X$}^{J_{1}}\times
\mbox{\boldmath$Y$}^{J_{2}}]^{J_{12}}\times\mbox{\boldmath$Z$}^{J_{3}}]^{J_{123}}= 
(-1)^{[J_1+J_2+J_{12}]+[J_{12}+J_3+J_{123}]+2J_{12}} \nonumber \\ 
\times  \sum_{J_{23}}
\left\{ \begin{array}{ccc} J_1 & J_2 & J_{12} \\
J_3 & J_{123} & J_{23}\end{array} \right\}^{\!_S}
[\mbox{\boldmath$X$}^{J_1}\times
[\mbox{\boldmath$Y$}^{J_2}\times\mbox{\boldmath$Z$}^{J_3}]^{J_{23}}]^{J_{123}}. 
\end{eqnarray}
For osp(1$|$2) we remind that the ``perimeter" $J_1+J_2+J_{12}$, related to a triangle, $\{J_1 J_2 J_{12}\}$ for example, may be integral or half-integral. Thus, summation indices, like $J_{12}$ or $J_{23}$, {\em implicitly} run by step of $\frac{1}{2}$.
(For su(2), the ``perimeters" are integral).

                    \section{Definition of iterated tensor operators $\mbox{\boldmath$S$}^{\kappa}$, from $\mbox{\boldmath$S$}^{\frac{1}{2}}$,
for osp(1$|$2)}\label{TensorIter}
\renewcommand{\theequation}{\ref{TensorIter}.\arabic{equation}}
\setcounter{equation}{0}

{\tt Three defining equations are necessary, namely}
\begin{equation}   \label{eq:STenspropertyfromDef}
[\mbox{\boldmath$S$}^{\frac{1}{2}}\times\mbox{\boldmath$S$}^{\frac{1}{2}}]^{0}=c_{0}\mbox{\boldmath$1$}^{0},
\mbox{\hspace{0.5em}}
[\mbox{\boldmath$S$}^{\frac{1}{2}}\times\mbox{\boldmath$S$}^{\frac{1}{2}}]^{\frac{1}{2}}=d_{0}\sqrt{2}\mbox{\boldmath$S$}^{\frac{1}{2}}
\mbox{\hspace{0.5em}and\hspace{0.5em}}[\mbox{\boldmath$S$}^{\frac{1}{2}}\times\mbox{\boldmath$S$}^{\kappa}]^{\kappa+\frac{1}{2}}=
\mbox{\boldmath$S$}^{\kappa+\frac{1}{2}}. 
\end{equation}
Here also, it is assumed that numbers $c_{0}, d_{0}\;\epsilon \; {\cal R}$.\par
As for su(2), it can be established that:
\begin{equation}   \label{eq:Sosp12withrank0andSymkhalf}
[\mbox{\boldmath$S$}^{\frac{1}{2}}\times\mbox{\boldmath$S$}^{\kappa}]^{\kappa+\frac{1}{2}}=
[\mbox{\boldmath$S$}^{\kappa}\times\mbox{\boldmath$S$}^{\frac{1}{2}}]^{\kappa+\frac{1}{2}}=
\mbox{\boldmath$S$}^{\kappa+\frac{1}{2}}
\mbox{\hspace{0.5em}and\hspace{0.5em}} \mbox{\boldmath$S$}^{0}=\mbox{\boldmath$1$}^{0}, 
\end{equation}
\begin{equation}   \label{eq:DefCoeffalpha}
[\mbox{\boldmath$S$}^{\frac{1}{2}}\times\mbox{\boldmath$S$}^{\kappa}]^{\kappa}=
[\mbox{\boldmath$S$}^{\kappa}\times\mbox{\boldmath$S$}^{\frac{1}{2}}]^{\kappa}=\alpha_{\kappa}\mbox{\boldmath$S$}^{\kappa},
\end{equation}
\begin{equation}   \label{eq:DefCoeffgamma}
[\mbox{\boldmath$S$}^{\frac{1}{2}}\times\mbox{\boldmath$S$}^{\kappa}]^{\kappa-\frac{1}{2}}=
[\mbox{\boldmath$S$}^{\kappa}\times\mbox{\boldmath$S$}^{\frac{1}{2}}]^{\kappa-\frac{1}{2}}=
\gamma_{\kappa}\mbox{\boldmath$S$}^{\kappa-\frac{1}{2}}.
\end{equation}
Instead of only one coefficient $\gamma_{k}$ for su(2), we have two coefficients
$\alpha_{\kappa}$ and $\gamma_{\kappa}$ to be determined afterwards.\par
In the same way, since 
$[\mbox{\boldmath$S$}^{\frac{1}{2}}\times\mbox{\boldmath$S$}^{\kappa}]^{\kappa''}=
[\mbox{\boldmath$S$}^{\kappa}\times\mbox{\boldmath$S$}^{\frac{1}{2}}]^{\kappa''}$,
we deduce the following property:
\begin{equation}   \label{eq:symosp12SkSkprime}
[\mbox{\boldmath$S$}^{\kappa}\times \mbox{\boldmath$S$}^{\kappa'}]^{\kappa''}=
[\mbox{\boldmath$S$}^{\kappa'}\times \mbox{\boldmath$S$}^{\kappa}]^{\kappa''}. 
\end{equation}
                                          \subsection{Definition of a binary parameter $\tau$}\label{BinaryTauDef}
A convenient parameter $\tau$, expressed by means of the integral part of any spin $\kappa$,
will be used for defining a  ``$\tau$-parity" of $\kappa$.
Our formulation is the following:
\begin{equation}   \label{eq:taudef}
\tau_{\kappa}=[\!\begin{array}{c}\kappa+\frac{1}{2}\end{array}\!]-[\kappa]=
\left\{\begin{array}{lc} 0 & \mbox{\hspace{0.5em} if $\kappa$ integral} \\ 1 & \mbox{\hspace{0.5em} if $\kappa$ half-integral} \end{array}\right..
\end{equation}
It looks like a kind of creator/annihilator for ``boson-fermion" with obvious properties:
\begin{equation}   \label{eq:tauproperties}
{\tau_{\kappa}}^{2}=\tau_{\kappa}, \mbox{\hspace{0.5em}} \tau_{\kappa}\tau_{\kappa-\frac{1}{2}}=0,
\mbox{\hspace{0.5em}} 
1-\tau_{\kappa-\frac{1}{2}}=\tau_{\kappa} 
\mbox{\hspace{0.5em}and} \mbox{\hspace{0.5em}}\tau_{\lambda+\kappa}=\tau_{\lambda}+\tau_{\kappa}-2\tau_{\lambda}\tau_{\kappa}.     
\end{equation}
Also, it can be used for representing any phase factor in the following way:
\begin{equation}   \label{eq:PhaseFactor}
(-1)^{2\kappa}=1- 2\tau_{\kappa}.
\end{equation}

                                  \subsection{Detailed calculation of $\alpha_{\kappa}$}\label{alphacalcul}
Consider the following product:
\begin{equation}   \label{eq:alphaCalc1}
[\mbox{\boldmath$S$}^{\frac{1}{2}}\times\mbox{\boldmath$S$}^{\kappa}]^{\kappa}=
[\mbox{\boldmath$S$}^{\frac{1}{2}}\times [\mbox{\boldmath$S$}^{\kappa -\frac{1}{2}}\times\mbox{\boldmath$S$}^{\frac{1}{2}}]^{\kappa}]^{\kappa}.
\end{equation}
A left-recoupling, use of formulas for $6$-$j^{S}$ with one spin $\frac{1}{2}$ and definition (\ref{eq:DefCoeffalpha}) lead to
\begin{equation}   \label{eq:Recurralpha}
\left( 2\kappa+(-1)^{2\kappa}\right) \alpha_{\kappa}=-\alpha_{\kappa -\frac{1}{2}}\sqrt{(2\kappa +1)(2\kappa -1)}. 
\end{equation}
Taken into account the initial condition $\alpha_{\frac{1}{2}}=d_{0}\sqrt{2}$, the solution is found to be:
\begin{equation}   \label{eq:Formulalphakappa}
\alpha_{\kappa}=(-1)^{2\kappa+1}\frac{(2\kappa+\tau_{\kappa})}{\sqrt{2\kappa(2\kappa+1)}} d_{0}.
\end{equation}
Alternative expressions useful for some further calculations are
\begin{equation}   \label{eq:Formulalphakappa2}
\left( \frac{(-1)^{2\kappa+1}\alpha_{\kappa}}{\sqrt{2\kappa(2\kappa+1)}}\right)=
\frac{(2\kappa+\tau_{\kappa})}{2\kappa(2\kappa+1)}d_0
=\left( \frac{\tau_{\kappa}}{2\kappa}\oplus\frac{\tau_{\kappa+\frac{1}{2}}}{2\kappa+1}\right)d_0.
\end{equation}

                                 \subsection{Detailed calculation of $\gamma_{\kappa}$}\label{gammacalcul}
We follow the same process as for the calculation of $\alpha_{\kappa}$, but in considering this time the product
\begin{equation}   \label{eq:gammaCalc1}
[\mbox{\boldmath$S$}^{\frac{1}{2}}\times\mbox{\boldmath$S$}^{\kappa}]^{\kappa-\frac{1}{2}}=
[\mbox{\boldmath$S$}^{\frac{1}{2}}\times [\mbox{\boldmath$S$}^{\kappa -\frac{1}{2}}\times\mbox{\boldmath$S$}^{\frac{1}{2}}]^{\kappa}]
^{\kappa-\frac{1}{2}}.
\end{equation}
Then the resulting recursion relation for $\gamma_{\kappa}$ is given by
\begin{equation}   \label{eq:gammarecurr}
\gamma_{\kappa}\left( 1- \frac{1}{2\kappa} \right) =\gamma_{\kappa-\frac{1}{2}} + \frac{(-1)^{2\kappa}}{2\kappa} {\alpha_{\kappa-\frac{1}{2}}}^{2}, 
\end{equation}
whose the solution is
\begin{equation}   \label{eq:GammaalphaRel}
\gamma_{\kappa}=\left( 2\kappa +(-1)^{2\kappa} \right) {\alpha_{\kappa}}^{2} +2\kappa c_{0}.  
\end{equation}
From expression (\ref{eq:Formulalphakappa}) the final form for $\gamma$ may be written as: 
\begin{equation}   \label{eq:Formulagammakappa1}
\gamma_{\kappa}=2\kappa(c_{0}+{d_{0}}^{2})-\frac{\tau_{\kappa}}{2\kappa} {d_{0}}^{2}, 
\end{equation}
or, under an alternative form, which separates $c_{0}$ from $d_{0}$:  
\begin{equation}   \label{eq:Formulagammakappa2}
\gamma_{\kappa}=2\kappa c_{0}+\frac{(2\kappa+\tau_{\kappa})(2\kappa-\tau_{\kappa})}{2\kappa} {d_{0}}^{2}. 
\end{equation}
Another interesting relationships between $\alpha$ and $\gamma$ may be found also in Appendix~A,
related to a study of special
reduced matrix elements. First values of $\gamma$ are the following:
\begin{equation}   \label{eq:FirstGamma}
\gamma_{0}=0, \mbox{\hspace{0.5em}}\gamma_{\frac{1}{2}}=c_{0}\mbox{\hspace{0.5em} and \hspace{0.5em}} 
\gamma_{1}=2\left( c_{0}+{d_{0}}^{2} \right). 
\end{equation}
\hspace*{1.5em}Now, given $\lambda$ and $\kappa$, the tensor products to be studied have ranks which vary by step of $\frac{1}{2}$ from
$\lambda +\kappa$ (the highest) to $|\kappa -\lambda|$ (the lowest). For the needs of our method of calculation, they must be organized in two sets ${\cal A}$ and ${\cal B}$ as follows:

\begin{eqnarray}   \label{eq:setA}
&& \mbox{\underline{Set ${\cal A}$}:\hspace{0.5em}}     
[\mbox{\boldmath$S$}^{\lambda}\times \mbox{\boldmath$S$}^{\kappa}]^{\lambda+\kappa} \cdots \cdots 
[\mbox{\boldmath$S$}^{\lambda}\times \mbox{\boldmath$S$}^{\kappa}]^{\sup (\lambda, \kappa)} \nonumber \\
&& \equiv \left\{ [\mbox{\boldmath$S$}^{\lambda}\times \mbox{\boldmath$S$}^{\kappa}]^{\lambda+\kappa-\pi}
\; ;\pi\,\epsilon\begin{array}{c}[0,\inf(\lambda,\kappa)]\end{array} \right\},
\end{eqnarray}
and
\begin{eqnarray}   \label{eq:setB}
&& \mbox{\underline{Set ${\cal B}$}:\hspace{0.5em}} 
[\mbox{\boldmath$S$}^{\lambda}\times \mbox{\boldmath$S$}^{\kappa}]^{\sup (\lambda, \kappa)-\frac{1}{2}} \cdots \cdots 
[\mbox{\boldmath$S$}^{\lambda}\times \mbox{\boldmath$S$}^{\kappa}]^{|\kappa-\lambda|} \nonumber \\
&& \equiv \left\{ [\mbox{\boldmath$S$}^{\lambda}\times \mbox{\boldmath$S$}^{\kappa}]^{|\lambda-\kappa|+\pi}
\; ;\pi\,\epsilon\begin{array}{c}[0,\inf(\lambda,\kappa)-\frac{1}{2}]; \inf(\lambda,\kappa)\geq \frac{1}{2}\end{array} \right\}.  
\end{eqnarray}

For the set ${\cal A}$, successive formulas will be obtained by starting from the highest rank $\lambda +\kappa$ (case $\pi=0$). For
${\cal B}$ one will start from the lowest rank $|\kappa-\lambda|$ (case $\pi=0$). 

                                           \section{Results for tensor products of highest ranks}\label{FirstTensFormulasH}
\renewcommand{\theequation}{\ref{FirstTensFormulasH}.\arabic{equation}}
\setcounter{equation}{0}

\hspace*{1.5em}The way to derive the formulas for each of the tensor operators
studied below is {\em invariable}.
It leads always to a recursion relation (to be solved) depending on the $\tau$-parities of $\lambda$ and $\kappa$.
                                \subsection{Presentation of the methods used for the calculation}\label{MethodIter}
The sequences are the following:\par
                  \hspace*{1.5em} (i) Right-recoupling:
\begin{eqnarray}   \label{eq:Seq1}
[\mbox{\boldmath$S$}^{\lambda}\times \mbox{\boldmath$S$}^{\kappa}]^{\lambda+\kappa-\pi}=
[[\mbox{\boldmath$S$}^{\lambda-\frac{1}{2}}\times\mbox{\boldmath$S$}^{\frac{1}{2}}]^{\lambda} \times
\mbox{\boldmath$S$}^{\kappa}]^{\lambda+\kappa-\pi} = (-1)^{2\lambda+2\kappa -2\pi +[\pi]}  \nonumber \\
\times  \sum_{J_{23}}
\left\{ \begin{array}{ccc} \lambda-\frac{1}{2} & \frac{1}{2} & \lambda \\
\kappa & \lambda+\kappa-\pi & J_{23}\end{array} \right\}^{\!_S}
[\mbox{\boldmath$S$}^{\lambda-\frac{1}{2}}\times
[\mbox{\boldmath$S$}^{\frac{1}{2}}\times\mbox{\boldmath$S$}^{\kappa}]^{J_{23}}]^{\lambda+\kappa-\pi}.
\end{eqnarray}
Except for the cases $\pi=0$ or $\pi=\frac{1}{2}$, $J_{23}$ takes three values, namely $\kappa-\frac{1}{2}, 
\kappa, \kappa+\frac{1}{2}$.
Evaluation of $6$-$j^{S}$ symbols thanks to the tables which can be found in ref. \cite{L.B.Nuovo.I}, as well as the use
of eqs.~(\ref{eq:DefCoeffalpha})-(\ref{eq:DefCoeffgamma}) 
result in a recursion relation like
\begin{eqnarray}   \label{eq:ReccurSHigh}
\lefteqn{ 
[\mbox{\boldmath$S$}^{\lambda}\times \mbox{\boldmath$S$}^{\kappa}]^{\lambda+\kappa-\pi}=
(\cdots) \gamma_{\kappa}
[\mbox{\boldmath$S$}^{\lambda-\frac{1}{2}}\times \mbox{\boldmath$S$}^{\kappa-\frac{1}{2}}]^{\lambda+\kappa-\pi}  } \nonumber \\
& & +(\cdots) \alpha_{\kappa}
[\mbox{\boldmath$S$}^{\lambda-\frac{1}{2}}\times \mbox{\boldmath$S$}^{\kappa}]^{\lambda+\kappa-\pi} 
+ (\cdots)
[\mbox{\boldmath$S$}^{\lambda-\frac{1}{2}}\times \mbox{\boldmath$S$}^{\kappa+\frac{1}{2}}]^{\lambda+\kappa-\pi},
\end{eqnarray}
where here we have considered $\kappa=\sup(\lambda, \kappa)$.\par
The analytical form taken by this recursion relation is different according to the $\tau$-parity of $\pi$, which can be integral or half-integral.
This comes from the involved $6$-$j^{S}$ symbols and their analytical formulas. \par \noindent
                 \hspace*{1.5em} (ii) Extraction of all coefficients in square roots as frontal common factors.\par
                 \hspace*{1.5em} (iii) Writing of {\em four} equations according to the $\tau$-parities of $\lambda, \kappa$:\par
We define four parity cases, denoted $(a), (b), (c), (d)$, according to the following table:
\begin{equation}   \label{eq:NotationsParities}
\begin{array}{|c|c|c|}
\lambda & \kappa & \mbox{notation} \\ \hline
\mbox{integral} & \mbox{integral} & (a) \\ \hline
\mbox{half-integral} & \mbox{half-integral} & (b) \\ \hline
\mbox{half-integral} & \mbox{integral} & (c) \\ \hline
\mbox{integral} & \mbox{half-integral} & (d) \\ \hline
\end{array}
\end{equation}
\par
                  \hspace*{1.5em} (iv) Summation of four finite series:\par
This can be done by using each time two of the four preceding equations and also results previously obtained for $\pi-\frac{1}{2}$ and 
$\pi-1$. Thus, the expected closure relation for the set ${\cal A}$, written as 
\begin{equation}   \label{eq:Final ResSHigh}
[\mbox{\boldmath$S$}^{\lambda}\times \mbox{\boldmath$S$}^{\kappa}]^{\lambda+\kappa-\pi}=
(\cdots) \mbox{\boldmath$S$}^{\lambda+\kappa-\pi},
\end{equation}
shows four different analytical coefficients $(\cdots)$ depending on the cases $(a)$, $(b)$, $(c)$, $(d)$.\par
                  \hspace*{1.5em} (v) Final result as a fully symmetrical expression in $\lambda, \kappa$:\par
{\em Symmetry $(\lambda\leftrightarrow\kappa)$ in formulas should be apparent, as
a consequence of the general property}~(\ref{eq:symosp12SkSkprime}).
However, this last step is perhaps the most difficult, because it's more or less heuristic.
The task consists in finding a single formula unifying the four cases.\par
\hspace*{1.5em}Let us illustrate the matter with the case where $\pi=\frac{1}{2}$.
At the end of step (iv) and with the  result of finite series summations, we are faced with four different terms we denote 
${\cal F}^{\frac{1}{2}}(\lambda,\kappa)$, not apparently symmetrical in $\lambda, \kappa$. One finds:\par
\[             \begin{array}{c|c|c|c|c|}
& (a) & (b) & (c) & (d) \\ \hline 
{\cal F}^{\frac{1}{2}}(\lambda,\kappa) & 0 & 2\lambda+2\kappa &  2\kappa &  2\lambda \\ \hline 
\end{array}        \]
The simplest way found for unifying these four terms is the following expression:
\begin{equation}   \label{eq:F12lambdakappa}
{\cal F}^{\frac{1}{2}}(\lambda,\kappa)=2\lambda \tau_{\kappa} + 2\kappa \tau_{\lambda},
\end{equation}
where binary factor $\tau$ has been defined in sect. \ref{BinaryTauDef}. 
It should be clear that the difficulties quickly increase for the cases $\pi \geq 1$ for obtaining fully symmetrical expressions
in $\lambda, \kappa$. \par
Let us now list the first formulas we have been able to obtain under the desired form.
\begin{equation}   \label{eq:Casepi0}
[\mbox{\boldmath$S$}^{\lambda}\times \mbox{\boldmath$S$}^{\kappa}]^{\lambda+\kappa}=
\mbox{\boldmath$S$}^{\lambda+\kappa}.
\end{equation}
\begin{equation}   \label{eq:Casepionehalf}
[\mbox{\boldmath$S$}^{\lambda}\times \mbox{\boldmath$S$}^{\kappa}]^{\lambda+\kappa-\frac{1}{2}}=
(-1)^{2\lambda+2\kappa} \frac{(2\lambda \tau_{\kappa}+2\kappa \tau_{\lambda})d_{0}}{\sqrt{(2\lambda)(2\kappa)(2\lambda+2\kappa)}}
\mbox{\boldmath$S$}^{\lambda+\kappa-\frac{1}{2}}.
\end{equation}
                                                \newcounter{property} \setcounter{property}{1}
                                                \renewcommand{\theequation}{\ref{FirstTensFormulasH}.\arabic{equation}.\theproperty}
                                                \addtocounter{equation}{-1}
This result implies the existence of  a non trivial set of zero tensor operators.\par
\hspace*{1.5em}{\underline{\em {Theorem 1}:}} 
From iterated tensor operators $\mbox{\boldmath$S$}^{k}$ of integral ranks, it can be built a zero tensor operator
by the following product:
\begin{equation}   \label{eq:ZeroTensOp}
[\mbox{\boldmath$S$}^{k}\times \mbox{\boldmath$S$}^{k'}]^{k+k'-\frac{1}{2}}=0, \mbox{\hspace{1em}} \forall k, k' \mbox{ integral}. 
\end{equation}
\renewcommand{\theequation}{\ref{FirstTensFormulasH}.\arabic{equation}} \setcounter{property}{0}
\begin{equation}   \label{eq:Casepi1}
[\mbox{\boldmath$S$}^{\lambda}\times \mbox{\boldmath$S$}^{\kappa}]^{\lambda+\kappa-1}=\frac{
2\lambda\cdot 2\kappa\cdot (2\lambda+2\kappa -1)(c_{0}+{d_{0}}^{2})-\tau_{\lambda}\tau_{\kappa}{d_{0}}^{2}}
{\sqrt{(2\lambda)(2\kappa)(2\lambda+2\kappa -1)}}\mbox{\boldmath$S$}^{\lambda+\kappa-1}.
\end{equation}
\begin{eqnarray}   \label{eq:Casepithreehalf}
\lefteqn{
[\mbox{\boldmath$S$}^{\lambda}\times \mbox{\boldmath$S$}^{\kappa}]^{\lambda+\kappa-\frac{3}{2}}=(-1)^{2\lambda+2\kappa} 
d_{0}(c_{0}+{d_{0}}^{2}) } \nonumber \\ [0.5em] 
& & \times \frac
{\scriptstyle(2\lambda -\tau_{\lambda})(2\kappa -\tau_{\kappa})(2\lambda+2\kappa-2+\tau_{\lambda}+\tau_{\kappa})
(2\lambda\tau_{\kappa}+2\kappa\tau_{\lambda}-1-\tau_{\lambda}-\tau_{\kappa}+\tau_{\lambda}\tau_{\kappa})}
{\sqrt{2\lambda(2\lambda -1)2\kappa(2\kappa -1)(2\lambda+2\kappa-1)(2\lambda+2\kappa-2)}}
\mbox{\boldmath$S$}^{\lambda+\kappa-\frac{3}{2}}. 
\end{eqnarray}

\begin{eqnarray} \label{eq:Casepi2}   
[\mbox{\boldmath$S$}^{\lambda}\times \mbox{\boldmath$S$}^{\kappa}]^{\lambda+\kappa-2}=\frac{1}{2}
{\scriptstyle(2\lambda -\tau_{\lambda})
(2\kappa -\tau_{\kappa})(2\lambda+2\kappa-2-\tau_{\lambda}-\tau_{\kappa}+2\tau_{\lambda}\tau_{\kappa})} 
(c_{0}+{d_{0}}^{2}) \mbox{\hspace{4em}}\nonumber \\ [0.5em]
\times \frac{ 
{\scriptstyle(2\lambda -1+\tau_{\lambda})(2\kappa -1+\tau_{\kappa})
(2\lambda+2\kappa-3+\tau_ {\lambda}+\tau_ {\kappa}-2\tau_{\lambda}\tau_{\kappa})}(c_{0}+{d_{0}}^{2})
{\scriptstyle -(1+2\tau_{\lambda}\tau_{\kappa})} {d_{0}}^{2} }
{ \sqrt{ \frac{ 2\lambda(2\lambda -1)2\kappa(2\kappa -1)(2\lambda+2\kappa-2)(2\lambda+2\kappa-3)} {2} }}
\mbox{\boldmath$S$}^{\lambda+\kappa-2}. \nonumber \\ 
\end{eqnarray}

\begin{eqnarray}   \label{eq:Casepifivehalf}
\lefteqn{
[\mbox{\boldmath$S$}^{\lambda}\times \mbox{\boldmath$S$}^{\kappa}]^{\lambda+\kappa-\frac{5}{2}}=(-1)^{2\lambda+2\kappa} 
d_{0}(c_{0}+{d_{0}}^{2}) }  \nonumber \\ [0.5em] 
& &   \begin{array}{c} \times \displaystyle{ \frac{1}{2} }  
{\scriptstyle(2\lambda -\tau_{\lambda})(2\kappa -\tau_{\kappa})(2\lambda +2\kappa -2-\tau_{\lambda}-\tau_{\kappa})
(2\lambda\tau_{\kappa}+2\kappa\tau_{\lambda}-2) } \\ [0.5em]
\times \left( 
\begin{array}{c}
\left( \begin{array}{c}{\scriptstyle(2\lambda-2-\tau_{\kappa}+\tau_{\lambda}\tau_{\kappa}) 
(2\kappa-2-\tau_{\lambda}+\tau_{\lambda}\tau_{\kappa})( 2\lambda+2\kappa-4+2\tau_{\lambda}+2\tau_{\kappa}-\tau_{\lambda}\tau_{\kappa}) } \\
+ {\scriptstyle 4(\tau_{\lambda}+\tau_{\kappa}-\tau_{\lambda}\tau_{\kappa})} \end{array} \right) {\displaystyle(c_{0}+{d_{0}}^{2})}   \\ [1.em]
- \scriptstyle{(\tau_{\lambda}+\tau_{\kappa}-\tau_{\lambda}\tau_{\kappa})} {\displaystyle {d_{0}}^{2} } \end{array}    \right)  \\ [1.8em]
\times {\displaystyle \frac{1} {\sqrt{ \frac{ 2\lambda(2\lambda -1)(2\lambda -2)2\kappa(2\kappa -1)(2\kappa -2)
(2\lambda+2\kappa-2)(2\lambda+2\kappa-3)(2\lambda+2\kappa-4)} {2} }} }
\mbox{\boldmath$S$}^{\lambda+\kappa-\frac{5}{2}}.\end{array}  \nonumber \\ 
\end{eqnarray}

\begin{eqnarray}   \label{eq:Casepithree}
\lefteqn{
[\mbox{\boldmath$S$}^{\lambda}\times \mbox{\boldmath$S$}^{\kappa}]^{\lambda+\kappa-3}= 
(c_{0}+{d_{0}}^{2}) \displaystyle{ \frac{1}{6} } 
{\scriptstyle(2\lambda -\tau_{\lambda})(2\kappa -\tau_{\kappa})(2\lambda +2\kappa -4-\tau_{\lambda}-\tau_{\kappa} +2\tau_\lambda\tau_\kappa)} } \nonumber  \\  
& &   \begin{array}{c} 
\times \left( 
\begin{array}{c}
\left( \begin{array}{c}{\scriptstyle(2\lambda-1+\tau_\lambda) (2\kappa-1+\tau_{\kappa})
( 2\lambda -2)(2\kappa -2)((2\lambda + 2\kappa -3) }  \\
\times  {\scriptstyle (2\lambda+2\kappa-5+\tau_{\lambda}+\tau_{\kappa}-2\tau_{\lambda}\tau_{\kappa}) } \end{array} \right) 
{\displaystyle \left( c_{0}+{d_{0}}^{2} \right)^{2}}   \\ [1.em]
- \scriptstyle{3 {\displaystyle{(}}(2\lambda-2)(2\kappa-2)(2\lambda+2\kappa-3)+2\tau_{\lambda}\tau_{\kappa} {\displaystyle{)}} } 
{\displaystyle  \left( {d_{0}}^{2}\right )\left( c_{0}+{d_{0}}^{2}\right) }   \\ [0.7em]
+ \scriptstyle {3\tau_{\lambda}\tau_{\kappa} }{\displaystyle \left( {d_{0}}^{2} \right)^{2} } \end{array}  \right)  \\ [1.4em]
\times {\displaystyle \frac{1} {\sqrt{ \frac{ 2\lambda(2\lambda -1)(2\lambda -2)2\kappa(2\kappa -1)(2\kappa -2)
(2\lambda+2\kappa-3)(2\lambda+2\kappa-4)(2\lambda+2\kappa-5)} {3\cdot 2} }} }
\mbox{\boldmath$S$}^{\lambda+\kappa-3}.  \end{array} \nonumber \\   
\end{eqnarray}

                                    \section{Closure relation for the set ${\cal A}$}\label{ClosureSetA}
\renewcommand{\theequation}{\ref{ClosureSetA}.\arabic{equation}}
\setcounter{equation}{0}

\hspace*{1.5em}As it follows from the formulas derived in sect.~\ref{FirstTensFormulasH}, we can assert that:
\begin{equation}   \label{eq:ClosRelosp12first}
[\mbox{\boldmath$S$}^{\lambda}\times \mbox{\boldmath$S$}^{\kappa}]^{\lambda+\kappa-\pi}=
\frac{(-1)^{4\pi(\lambda+\kappa)}\;
{\cal{P}}^{\pi}(\lambda,\kappa)}{E(\lambda,\kappa;\lambda+\kappa-\pi)
\bigtriangledown^{S}(\lambda\; \kappa \;\lambda+\kappa-\pi)}\; \mbox{\boldmath$S$}^{\lambda+\kappa-\pi},
\end{equation}
where the normalization factor $E$ (in square root) is the analog of the one found in the closure relation for su(2), see eq.~(\ref{eq:ClosureRelationsu2}), with a slight difference in the denominator, {\em i.e.}
\begin{equation}   \label{eq:normalizEosp12}
E(a,b;c)=\left[\frac{(2a)!(2b)!}{(2c)!}\right]^{\frac{1}{2}}.
\end{equation}
${\cal{P}}^{\pi}(\lambda,\kappa)$ is a polynomial symmetrical in $\lambda, \kappa$, namely
\begin{equation}   \label{eq:SymPpi}
{\cal{P}}^{\pi}(\lambda,\kappa)={\cal{P}}^{\pi}(\kappa,\lambda),
\end{equation}
Polynomials ${\cal{P}}^{\pi}$ satisfy (functional) recursion relations derived from eq.~(\ref{eq:ReccurSHigh}).  
Their thorough study is reported in Appendix~B.
They have the following expansion:
\begin{equation}   \label{eq:PolynP}
{\cal P}^{\,\pi}(\lambda,\kappa)=\left( {d_{0}} \right)^{\tau_{\pi}}
\sum_{m=0}^{m=[\pi]} x_{\,m}^{\,\pi}(\lambda,\kappa)
\left(c_{0}+{d_{0}}^{2}\right)^{[\pi]-m}({d_{0}}^{2})^{m},
\end{equation}
with the same symmetry property in $\lambda, \kappa$ for the coefficients $x$, namely
\begin{equation}   \label{eq:symmcoeffx}
x_{\,m}^{\,\pi}(\lambda,\kappa)=x_{\,m}^{\,\pi}(\kappa,\lambda).
\end{equation}
                                                   \hspace*{1.5em}{\underline{\em {Conjecture 1}:}}
Coefficients $x_{\,m}^{\,\pi}(\lambda,\kappa)$ are integral numbers.\par 
This is a reasonable and weak conjecture agreeing with the results obtained for the lowest values of $\pi$. \par 
Having in mind that, of course, any tensor operator $\mbox{\boldmath$S$}^{\kappa}$ depends implicitly on  $c_{0}, d_{0}$,  
we may write generally:
\begin{eqnarray}   \label{eq:ClosRelosp12SetA}
[\mbox{\boldmath$S$}^{a}\times \mbox{\boldmath$S$}^{b}]^{c}=
\frac{(-1)^{4(a+b)(a+b+c)} \;
{\cal{P}}^{a+b-c}(a,b)}{E(a,b;c)
\bigtriangledown^{S}(a\; b \;c)}\; \mbox{\boldmath$S$}^{c}, \;\forall \;c\; \epsilon \; [\sup(a,b),a+b].
\end{eqnarray}

                              \section{Results for tensor products of lowest ranks}\label{FirstTensFormulasL}
\renewcommand{\theequation}{\ref{FirstTensFormulasL}.\arabic{equation}}
\setcounter{equation}{0}
In this section, examples will be given by assuming $\kappa=\sup(\lambda, \kappa)$.\par
The methods used for deriving formulas for $[\mbox{\boldmath$S$}^{\lambda}\times \mbox{\boldmath$S$}^{\kappa}]^{\kappa-\lambda+\pi}$
are similar to those followed in sect.~\ref{FirstTensFormulasH}, with obvious slight changes. For instance step (i) becomes
\begin{eqnarray}   \label{eq:Seq1prime}
[\mbox{\boldmath$S$}^{\lambda}\times \mbox{\boldmath$S$}^{\kappa}]^{\kappa-\lambda+\pi}=
[[\mbox{\boldmath$S$}^{\lambda-\frac{1}{2}}\times\mbox{\boldmath$S$}^{\frac{1}{2}}]^{\lambda} \times
\mbox{\boldmath$S$}^{\kappa}]^{\kappa-\lambda+\pi} = (-1)^{2\kappa+[\pi]} \nonumber \\
\times  \sum_{J_{23}}
\left\{ \begin{array}{ccc} \lambda-\frac{1}{2} & \frac{1}{2} & \lambda \\
\kappa & \kappa-\lambda+\pi & J_{23}\end{array} \right\}^{\!_S}
[\mbox{\boldmath$S$}^{\lambda-\frac{1}{2}}\times
[\mbox{\boldmath$S$}^{\frac{1}{2}}\times\mbox{\boldmath$S$}^{\kappa}]^{J_{23}}]^{\kappa-\lambda+\pi},
\end{eqnarray} 
and the recursion relation between tensor operators reads now:
\begin{eqnarray}   \label{eq:ReccurSLow}
\lefteqn{ 
[\mbox{\boldmath$S$}^{\lambda}\times \mbox{\boldmath$S$}^{\kappa}]^{\kappa-\lambda+\pi}=
(\cdots) \gamma_{\kappa}
[\mbox{\boldmath$S$}^{\lambda-\frac{1}{2}}\times \mbox{\boldmath$S$}^{\kappa-\frac{1}{2}}]^{\kappa-\lambda+\pi}  } \nonumber \\
& & +(\cdots) \alpha_{\kappa}
[\mbox{\boldmath$S$}^{\lambda-\frac{1}{2}}\times \mbox{\boldmath$S$}^{\kappa}]^{\kappa-\lambda+\pi}
+ (\cdots)
[\mbox{\boldmath$S$}^{\lambda-\frac{1}{2}}\times \mbox{\boldmath$S$}^{\kappa+\frac{1}{2}}]^{\kappa-\lambda+\pi}.
\end{eqnarray}
Steps (ii), (iii) and (iv) remain formally identical. However step (v) cannot include a symmetry such as $\lambda \leftrightarrow \kappa$, but this
very important point will become clearer below, exhibiting a fundamental variable change like ($\lambda \rightarrow \lambda, \;  
\kappa \rightarrow |\kappa-\lambda|+\pi$), involving in addition a product of $\gamma$ coefficients.\par
$\bullet$ Let us start with the simplest case where $\pi=0$.

\begin{equation}   \label{eq:caseLowpi0}
[\mbox{\boldmath$S$}^{\lambda}\times \mbox{\boldmath$S$}^{\kappa}]^{\kappa -\lambda}=
( \gamma_{\kappa}\cdots \gamma_{\kappa-\lambda+\frac{1}{2}} )
\mbox{\boldmath$S$}^{\kappa -\lambda}. 
\end{equation}
An alternative expression of (\ref{eq:caseLowpi0}) may be written with familiar normalization factors like those occuring in denominator of
eq.~(\ref{eq:ClosRelosp12SetA}), however with a change in the order of spins in $E$, and this gives an idea for the future hypothesis to come:
\begin{equation}   \label{eq:caseLowpi0bis}
[\mbox{\boldmath$S$}^{\lambda}\times \mbox{\boldmath$S$}^{\kappa}]^{\kappa -\lambda}=
\frac{( \gamma_{\kappa}\cdots \gamma_{\kappa-\lambda+\frac{1}{2}} )}{E(\lambda,\kappa-\lambda;\kappa)\bigtriangledown^{S}(\lambda\,\kappa\;\kappa-\lambda)}
\mbox{\boldmath$S$}^{\kappa -\lambda}. 
\end{equation}
A particular case is the {\tt \small \bf scalar product} \rm 
\footnote{The trivial case $[\mbox{\boldmath$S$}^{0}\times \mbox{\boldmath$S$}^{0}]^{0}=\mbox{\boldmath$S$}^{0}$ is found in the set ${\cal A}$
with $\lambda=\kappa=\pi=0$.} 
, namely
\begin{equation}   \label{eq:ScalarProductL}
[\mbox{\boldmath$S$}^{\kappa}\times \mbox{\boldmath$S$}^{\kappa}]^{0}=
( \gamma_{\kappa}\cdots \gamma_{\frac{1}{2}} ) \mbox{\boldmath$S$}^{0} \equiv
 \frac{( \gamma_{\kappa}\cdots \gamma_{\frac{1}{2}} )}{{E(\kappa,0;\kappa)\bigtriangledown^{S}(\kappa\,\kappa\;0)}}
\mbox{\boldmath$S$}^{0}, \mbox{\hspace{0.5em}} \forall \kappa \geq \frac{1}{2 }.  
\end{equation}
$\bullet$ Our result for the next case, $\pi=\frac{1}{2}$, is the following:
\begin{eqnarray}   \label{eq:caseLowpi1/2}
\lefteqn{
[\mbox{\boldmath$S$}^{\lambda}\times \mbox{\boldmath$S$}^{\kappa}]^{\kappa -\lambda+\frac{1}{2}}= (-1)^{2\kappa+1} \;
( \gamma_{\kappa}\cdots \gamma_{\kappa-\lambda+1} ) }  \nonumber \\ 
& &  \times
\frac{\left(2\lambda \tau_{\kappa-\lambda+\frac{1}{2}}+2(\kappa-\lambda+\frac{1}{2})\tau_{\lambda}\right)d_{0}}
{E(\lambda,\kappa-\lambda+\frac{1}{2};\kappa)\bigtriangledown^{S}(\lambda\,\kappa\;\kappa-\lambda+\frac{1}{2})}
\mbox{\boldmath$S$}^{\kappa -\lambda+\frac{1}{2}}.  
\end{eqnarray}
One recognizes the polynomial ${\cal P}^{\frac{1}{2}}$ encountered first in eq.~(\ref{eq:Casepionehalf}) and generally defined by 
eq.~(\ref{eq:ClosRelosp12first}) with the following change of variables $\lambda \rightarrow \lambda, \,
\kappa \rightarrow \kappa -\lambda+\frac{1}{2}$ . Therefore
\begin{eqnarray}   \label{eq:caseLowpi1/2bis}
\lefteqn{
[\mbox{\boldmath$S$}^{\lambda}\times \mbox{\boldmath$S$}^{\kappa}]^{\kappa -\lambda+\frac{1}{2}}=  (-1)^{2\kappa+1} \;
( \gamma_{\kappa}\cdots \gamma_{\kappa-\lambda+1} ) }  \nonumber \\ 
& & \times
\frac{ {{\cal{P}}^{\frac{1}{2}} (\lambda,\kappa -\lambda+\frac{1}{2})}}
{E(\lambda,\kappa-\lambda+\frac{1}{2};\kappa)\bigtriangledown^{S}(\lambda\,\kappa\;\kappa-\lambda+\frac{1}{2})}
\mbox{\boldmath$S$}^{\kappa -\lambda+\frac{1}{2}}.  
\end{eqnarray}
                            \setcounter{property}{1}
                            \renewcommand{\theequation}{\ref{FirstTensFormulasL}.\arabic{equation}.\theproperty}
                            \addtocounter{equation}{-1}
 
Exactly as for {\em Theorem 1} this result implies the existence of  a non trivial set of zero tensor operators. \\ [0.5em]\par
\hspace*{1.5em}{\underline{\em {Theorem 2}:}} 
From iterated tensor operators $\mbox{\boldmath$S$}^{k}$ with $k$  integral and $\mbox{\boldmath$S$}^{\kappa}$ with $\kappa$ half-integral, 
it can be built zero tensor operators by the following product:
\begin{equation}   \label{eq:ZeroTensOp2}
[\mbox{\boldmath$S$}^{k}\times \mbox{\boldmath$S$}^{\kappa}]^{|k-\kappa|+\frac{1}{2}}=0, 
{\mbox{\hspace{1em}} \forall k \mbox{ integer }, \mbox{ and \hspace{0.5em}} \forall \kappa \mbox{ half-integer}}. 
\end{equation}
\renewcommand{\theequation}{\ref{FirstTensFormulasL}.\arabic{equation}} \setcounter{property}{0}
As an example one can write:
\begin{equation}   \label{eq:ExmplZeroTensOp2}
[\mbox{\boldmath$S$}^{\lambda}\times \mbox{\boldmath$S$}^{\lambda+\frac{1}{2}+p}]^{p+1}=0, 
\mbox{\hspace{1em}} \forall \lambda
\mbox{ \hspace{0.1em} {\scriptsize integer or half-integer}},
\mbox{ and \hspace{0.2em}} \forall p \mbox{\hspace{0.2em} {\scriptsize integer}}. 
\end{equation}

                                           \section{Closure relation for the set ${\cal B}$}\label{ClosureSetB}
\renewcommand{\theequation}{\ref{ClosureSetB}.\arabic{equation}}
\setcounter{equation}{0}

\hspace*{1.5em}As it can be guessed from the preceding short study, 
one can assert the following formulation
($\kappa \geq \lambda$ is adopted for convenience):  
\begin{eqnarray}   \label{eq:ClosRelosp12second}
\lefteqn{
[\mbox{\boldmath$S$}^{\lambda}\times \mbox{\boldmath$S$}^{\kappa}]^{\kappa-\lambda+\pi}= (-1)^{4\pi(\kappa+\pi)} 
( \gamma_{\kappa} \cdots\gamma_{\kappa-\lambda+\pi+\frac{1}{2}} )                       } \nonumber \\
& & \times \frac{{\cal{Q}}^{\pi} (\lambda;\kappa)} {E(\lambda,\kappa-\lambda+\pi;\kappa)
\bigtriangledown^{S}(\lambda\; \kappa \;\kappa-\lambda+\pi)\; }\mbox{\boldmath$S$}^{\kappa-\lambda+\pi}, 
\end{eqnarray}
where ${\cal Q}^{\pi}(\lambda;\kappa)$ is a polynomial in $\lambda,\kappa$. Like polynomials ${\cal P}^{\pi}$, ${\cal Q}^{\pi}$
satisfy recursion relations issued from eq.~(\ref{eq:ReccurSLow}). They can be found listed in Appendix~B. \par
A glance at their respective recursion relations does not exhibit immediate similarities between these polynomials, though there is a special one
of importance.\par
However, thanks to a lengthy study we have carried out for obtaining explicit analytical formulas for tensor operators
(sets ${\cal A}$ and ${\cal B}$ are included),
involving the four parity cases $(a)$, $(b)$, $(c)$, $(d)$, from 
\[ [\mbox{\boldmath$S$}^{\frac{1}{2}} \times \mbox{\boldmath$S$}^{\kappa}]^{\kappa+\frac{1}{2}} \longrightarrow
[\mbox{\boldmath$S$}^{\frac{1}{2}} \times \mbox{\boldmath$S$}^{\kappa}]^{\kappa}
\longrightarrow
[\mbox{\boldmath$S$}^{\frac{1}{2}} \times \mbox{\boldmath$S$}^{\kappa}]^{\kappa-\frac{1}{2}}, \]
up to
\[ [\mbox{\boldmath$S$}^{\frac{7}{2}} \times \mbox{\boldmath$S$}^{\kappa}]^{\kappa+\frac{7}{2}}
\cdots\longrightarrow
[\mbox{\boldmath$S$}^{\frac{7}{2}} \times \mbox{\boldmath$S$}^{\kappa}]^{\kappa}
\cdots \longrightarrow[\mbox{\boldmath$S$}^{\frac{7}{2}} \times \mbox{\boldmath$S$}^{\kappa}]^{\kappa-\frac{7}{2}}, \]
we can {\em conjecture} that the following property holds:\par

\hspace*{1.5em}{\underline{\em {Conjecture 2}:}} The relation between ${\cal Q}^{\pi}$ and ${\cal P}^{\pi}$ polynomials is given by 
\begin{equation}   \label{eq:RelQPpolynomials}
{\cal Q}^{\pi}(\lambda;\kappa)={\cal P}^{\pi}(\lambda,\kappa - \lambda+\pi).
\end{equation}
From this property, after some variable changes, it follows that a closure relation can be explicitely written for the set ${\cal B}$:
\begin{equation}  \label{eq:ClosRelosp12SetB}
\begin{array}{lll}  
[\mbox{\boldmath$S$}^{a}\times \mbox{\boldmath$S$}^{b}]^{c}=&
\displaystyle \frac{(-1)^{4\left(\inf(a,b)+c\right)(a+b+c)} \;
{\cal{P}}^{c-|b-a|}\,\left(\inf(a,b),c\right)} {E\left(\inf(a,b),c;\sup(a,b)\right)  \bigtriangledown^{S}(a\; b \;c)}  
\gamma_{\,\sup(a,b)}\cdots \gamma_{c+\frac{1}{2}} \mbox{\boldmath$S$}^{c}, &  \\ [1em]
& \forall \;c\; \epsilon  [|b-a|,\sup(a,b)-\frac{1}{2}], \mbox{and } \inf(a,b)\geq \frac{1}{2}. &
\end{array} 
\end{equation}

                                           \section{Unified form of the closure relation for tensor operators $ \mbox{\boldmath$S$}^{\kappa}$}\label{UnifiedClosure}
\renewcommand{\theequation}{\ref{UnifiedClosure}.\arabic{equation}}
\setcounter{equation}{0}
In the preceding studies on sets ${\cal A}$ and ${\cal B}$, some interesting features can be noticed.\par
\begin{equation}   \label{eq:LimitsA}
\sup(a,b,c)=c \mbox{\hspace{0.5em} and} \mbox{\hspace{0.5em}} \inf(\sup(a,b),c)=\sup(a,b), \mbox{\hspace{0.5em} for ${\cal A}$}.  
\end{equation}
\begin{equation}   \label{eq:LimitsB}
\sup(a,b,c)=\sup(a,b) \mbox{\hspace{0.5em} and} \mbox{\hspace{0.5em}} \inf(\sup(a,b),c)=c, \mbox{\hspace{0.5em} for ${\cal B}$}.  
\end{equation}
We choose to adopt the following convention regarding the notation with dots for the ordered product of $\gamma$ coefficients, enclosed inside brackets like $(\cdots)$:
\begin{equation}   \label{eq:ConvGammas}
(\gamma_{c_{>}}\cdots\gamma_{c_{<}})=1\mbox{\hspace{0.5em} if  \hspace{0.3em} $c_{<} > c_{>}$ } (convention).
\end{equation}
For instance:
\begin{equation}   \label{eq:ExmplCOnvGamma}
(\gamma_{c}\cdots\gamma_{c+\frac{1}{2}})=1.
\end{equation}
Thus a single form of eqs.~(\ref{eq:ClosRelosp12SetA}), (\ref{eq:ClosRelosp12SetB}) can be written as a general {\em closure relation}:
\begin{eqnarray}   \label{eq:UnifCloseS}
[\mbox{\boldmath$S$}^{a}\times \mbox{\boldmath$S$}^{b}]^{c}=
(-1)^{\tau_{a+b+c}+2\sup(a,b,c)\tau_{a+b+c}} 
( \gamma_{\,\sup(a,b,c)} \cdots \gamma_{c+\frac{1}{2}} )    \nonumber \\
\times \frac{ 
{\cal P}^{\pi(a,b,c)}\bigl( \inf(a,b),\inf(\sup(a,b),c) \bigr) \bigtriangleup^{S}(a\; b \;c) \mbox{\boldmath$S$}^{c} }
{ E( \inf(a,b),\inf(\sup(a,b),c);\sup(a,b,c) ) }
,
\end{eqnarray}
where pseudo-degree $\pi$ initially defined by 
\begin{equation}   \label{eq:pidef}
\pi(a,b,c)=\inf(a,b)+\inf( \sup(a,b),c)-\sup(a,b,c),
\end{equation}
has a form fully symmetrical in $a,b,c$ given by
\begin{equation}   \label{eq:pidefprime}
\pi(a,b,c)=a+b+c-2\sup(a,b,c).
\end{equation}

                    \section{{\em Quasi}-final demonstration for a $\bigtriangleup^{S}$-sum rule related to osp(1$|$2)}\label{PrefinalProof}
\renewcommand{\theequation}{\ref{PrefinalProof}.\arabic{equation}}
\setcounter{equation}{0}
This is an analog of the one developed for su(2) in sect.~\ref{Finalsu2}.\par
 
Consider the triple product $[ [\mbox{\boldmath$S$}^{a}\times \mbox{\boldmath$S$}^{b}]^{c}\times\mbox{\boldmath$S$}^{d} ]^{e}$.
Eq.~(\ref{eq:UnifCloseS}) used twice leads to:
\begin{eqnarray}   \label{eq:osp12SaSbrkcSdrkdrke}
[ [\mbox{\boldmath$S$}^{a}\times \mbox{\boldmath$S$}^{b}]^{c}\times\mbox{\boldmath$S$}^{d} ]^{e}=
( \gamma_{\,\sup(a,b,c)} \cdots \gamma_{c+\frac{1}{2}} ) ( \gamma_{\,\sup(c,d,e)} \cdots \gamma_{e+\frac{1}{2}} )  \nonumber \\
\times (-1)^{\tau_{a+b+d+e}}(-1)^{2\sup(a,b,c)\tau_{a+b+c}+ 2\sup(c,d,e)\tau_{c+d+e}}  \nonumber \\
\times \frac{ {\cal P}^{\pi(a,b,c)}\bigl( \inf(a,b),\inf(\sup(a,b),c) \bigr) }
{ E( \inf(a,b),\inf(\sup(a,b),c);\sup(a,b,c) ) } \nonumber \\
\times \frac{ {\cal P}^{\pi(c,d,e)}\bigl( \inf(c,d),\inf(\sup(c,d),e) \bigr) } { E( \inf(c,d),\inf(\sup(c,d),e);\sup(c,d,e) )} \nonumber \\
\times  \bigtriangleup^{S}(a\, b \,c)\bigtriangleup^{S}(c\, d \,e) \mbox{\boldmath$S$}^{e}. 
\end{eqnarray}
Next, a right-recoupling is carried out and yields
\begin{eqnarray}   \label{eq:OSP12RightRecoupl}
[[\mbox{\boldmath$S$}^{a}\times
\mbox{\boldmath$S$}^{b}]^{c}\times\mbox{\boldmath$S$}^{d}]^{e}= 
(-1)^{[a+b+c]+[c+d+e]+2c}  \nonumber \\ 
\times  \sum_{f}
\left\{ \begin{array}{ccc} a & b & c  \\
d & e & f \end{array} \right\}^{\!_S}
[\mbox{\boldmath$S$}^{a}\times
[\mbox{\boldmath$S$}^{b}\times\mbox{\boldmath$S$}^{d}]^{f}]^{e}.
\end{eqnarray}
Closure relation~(\ref{eq:UnifCloseS}) then is used for $[\mbox{\boldmath$S$}^{b}\times\mbox{\boldmath$S$}^{d}]^{f}$
and $[\mbox{\boldmath$S$}^{a}\times\mbox{\boldmath$S$}^{f}]^{e}$.\par
Thus an alternative expression for 
$[[\mbox{\boldmath$S$}^{a}\times\mbox{\boldmath$S$}^{b}]^{c}\times\mbox{\boldmath$S$}^{d}]^{e}$ is obtained, namely

\begin{eqnarray}   \label{eq:OSP12RightRecouplClos}
[[\mbox{\boldmath$S$}^{a}\times
\mbox{\boldmath$S$}^{b}]^{c}\times\mbox{\boldmath$S$}^{d}]^{e}= 
(-1)^{[a+b+c]+[c+d+e]+2c}  \sum_{f} 
\left\{ \begin{array}{ccc} a & b & c  \\ d & e & f \end{array} \right\}^{\!_S} \nonumber \\ 
\times  (-1)^{2(a+b+d+e)}(-1)^{2\sup(b,d,f)\tau_{b+d+f }+ 2\sup(a,f,e)\tau_{a+f+e}} \nonumber \\
\times  ( \gamma_{\,\sup(b,d,f)} \cdots \gamma_{f+\frac{1}{2}} ) \frac{ {\cal P}^{\pi(b,d,f)}\bigl( \inf(b,d),\inf(\sup(b,d),f) \bigr) }
{ E( \inf(b,d),\inf(\sup(b,d),f);\sup(b,d,f) ) } \nonumber \\
\times ( \gamma_{\,\sup(a,f,e)} \cdots \gamma_{e+\frac{1}{2}} )
\frac{ {\cal P}^{\pi(a,f,e)}\bigl( \inf(a,f),\inf(\sup(a,f),e) \bigr) } 
{ E( \inf(a,f),\inf(\sup(a,f),e);\sup(a,f,e) )} \nonumber \\
\times \bigtriangleup^{S}(b\, d \,f)\bigtriangleup^{S}(a\, f \,e) \mbox{\boldmath$S$}^{e}. 
\end{eqnarray}
Coefficients $E$ occuring in eqs.~(\ref{eq:osp12SaSbrkcSdrkdrke}), (\ref{eq:OSP12RightRecouplClos}) may be written differently. For instance:
\begin{equation} \label{eq:ExpressCoeffE}
E( \inf(a,b),\inf(\sup(a,b),c);\sup(a,b,c) )=
\frac{\sqrt{(2a)!(2b)!(2c)!} } {(2\sup(a,b,c))!}.
\end{equation} 
Then, an identification of both aforementioned equations over $\mbox{\boldmath$S$}^{e}$ shows that {\em coefficients in square root} like
$\sqrt{(2a)!(2b)!(2d)!(2e)!}$ {\em vanish}.\par
\hspace{1.5em}Thus a {\tt preliminary form} of the $\bigtriangleup^{S}$-sum rule may be presented as follows:
\begin{eqnarray} \label{eq:PreliminDeltaSSumRule}
(-1)^{2\sup(a,b,c)\tau_{a+b+c} + 2\sup(c,d,e)\tau_{c+d+e}}   \nonumber \\
\times ( \gamma_{\,\sup(a,b,c)} \cdots \gamma_{c+\frac{1}{2}} ) {\cal P}^{\pi(a,b,c)}\bigl( \inf(a,b),\inf(\sup(a,b),c) \bigr) \nonumber \\
\times ( \gamma_{\,\sup(c,d,e)} \cdots \gamma_{e+\frac{1}{2}} ) {\cal P}^{\pi(c,d,e)}\bigl( \inf(c,d),\inf(\sup(c,d),e) \bigr) \nonumber \\
\times \begin{array}{c} \frac {(2\sup(a,b,c))!(2\sup(c,d,e))!} { (2c)!} \end{array}
\bigtriangleup^{S}(a\, b \,c)\bigtriangleup^{S}(c\, d \,e)  \nonumber \\
= (-1)^{[a+b+c]+[c+d+e]+2c}  \sum_{f} 
\left\{ \begin{array}{ccc} a & b & c  \\ d & e & f \end{array} \right\}^{\!_S} \nonumber \\
\times (-1)^{2\sup(b,d,f)\tau_{b+d+f} + 2\sup(a,f,e)\tau_{a+f+e}}   \nonumber \\
\times ( \gamma_{\,\sup(b,d,f)} \cdots \gamma_{f+\frac{1}{2}} ) {\cal P}^{\pi(b,d,f)}\bigl( \inf(b,d),\inf(\sup(b,d),f) \bigr) \nonumber \\
\times ( \gamma_{\,\sup(a,f,e)} \cdots \gamma_{e+\frac{1}{2}} ) {\cal P}^{\pi(a,f,e)}\bigl( \inf(a,f),\inf(\sup(a,f),e) \bigr) \nonumber \\
\times \begin{array}{c} \frac {(2\sup(b,d,f))!(2\sup(a,f,e))!} { (2f)!} \end{array}
\bigtriangleup^{S}(b\, d \,f)\bigtriangleup^{S}(a\, f \,e).  
\end{eqnarray}
 
                    \subsection{Invariance check with respect to the $6$-$j^{S}$ orthogonality relation}\label{CheckOrtho6jS}  
Consider the pseudo-orthogonality relation between $6$-$j^{S}$ symbols \cite{L.B.Nuovo.I}:
\begin{eqnarray}   \label{eq:Orthog6jS}
\sum_{x}(-1)^{[a+b+x]+[d+e+x]+2x}
\left\{ \begin{array}{ccc} a & b & x \\
d & e & f \end{array} \right\}^{\!_S}  
\left\{ \begin{array}{ccc} a & b & x \nonumber \\
d & e & f' \end{array} \right\}^{\!_S}  \\ 
= (-1)^{[a+e+f]+[b+d+f]+2f} \delta_{f,f'}. 
\end{eqnarray}
By operating both left/right hand-sides of eq.~(\ref{eq:PreliminDeltaSSumRule}) by 
$\displaystyle \sum_{c} 
\left\{ \begin{array}{ccc} a & b & c  \\ d & e & f' \end{array} \right\}^{\!_S}$, it can be checked that one retrieves the same analytical form of the $\bigtriangleup^{S}$-sum rule.

                    \section{Ultimate identification process over $(c_{0}+{d_{0}}^{2}), {d_{0}}^{2}$}\label{FinalProof}  
\renewcommand{\theequation}{\ref{FinalProof}.\arabic{equation}}
\setcounter{equation}{0}

{\tt Short warning about osp(1$|$2)-specific troubles to come:}\par
\hspace*{1.5em}
For su(2) or su$_q$(2), the identification process over tensor operators $\mbox{\boldmath$S$}^{e}$ was very simple and leads straightforwardly to the final desired result, because a single parameter, $c_0$, vanishes after simplification by 
$\left(c_0/\sqrt{2}\right)^{a+b+d-e}$ or $\left(c_0/\sqrt{[2]}\right)^{a+b+d-e}$.
\par
\hspace*{1.5em}
On the other hand, for osp(1$|2$), the situation becomes much more complex. Indeed, we deal with two independent parameters, namely $(c_{0}+{d_{0}}^{2})$ and ${d_{0}}^{2}$, or equivalently $c_0$ and ${d_{0}}^{2}$.
Furthermore, we will need to carry out a countable set of identifications with some $x(a,b,c,d,e;f)$ coefficients of binomials like
$(c_{0}+{d_{0}}^{2})^{\Omega-m}({d_{0}}^{2})^{m}$, related to homogeneous polynomials of degree $\Omega$.
\par
\hspace*{1.5em}
As it will be seen in the sequel, regarding summation indices (integral or half-integral), {\em a priori} it seems that we should expect only two kinds of osp(1$|2$) $\bigtriangleup^{S}$-sum rule for which our aim is to find an unified form. This could be reasonably accepted.
\par
\hspace*{1.5em}
{\it However this fact implies that deeply hidden identities between $x$ coefficients should exist. If this property does not hold (or cannot be proved), then we should also imagine an ``infinity" of $\bigtriangleup^{S}$-sum rule, which would force us to re-consider independence properties of the symbols $6$-$j^S$ as well as the status of  osp(1$|2$) itself, a thinking rather inconceivable.}
\par
\hspace*{1.5em}
All that summarizes our present state of problems which still need some enlightenment. Let us now give an idea of the detailed calculations to be carried out.  
\par 
{\tt Details on the polynomial analysis}:\par
\hspace*{1.5em}Consider the product of two polynomials ${\cal P}$. From eq.~(\ref{eq:FormofPolynP}) we obtain:
\begin{eqnarray}   \label{eq:ProdPabcPcde}
{\cal P}^{\,\omega}\bigl(\inf(a,b),\inf(\sup(a,b),c)\bigr) {\cal P}^{\,\omega'}\bigl(\inf(c,d),\inf(\sup(c,d),e)\bigr) \nonumber \\
= {d_0}^{\tau_{a+b+c}+\tau_{c+d+e}} {\cal R}^{\Omega}(a,b;c|c,d,e),  
\end{eqnarray}
where ${\cal R}^{\Omega}$ is an homogeneous polynomial in $(c_{0}+{d_{0}}^{2}), {d_{0}}^{2}$,
of degree $\Omega=[\omega]+[\omega']$:  
\begin{equation}   \label{eq:DefPolynR}
{\cal R}^{\Omega}(a,b;c|c,d;e))=
\sum_{m=0}^{m=\Omega} y_{m}^{\omega,\omega'}(a,b;c| c,d;e) {(c_{0}+{d_{0}}^{2})}^{\Omega -m} \left({d_{0}}^{2}\right)^{m}.
\end{equation}
Coefficients $y_{m}^{\omega,\omega'}$ are given as function of coefficients $x$ defined in eq.~(\ref{eq:FormofPolynP}):
\begin{eqnarray}   \label{eq:DEfCoeffy}
\lefteqn{ y_{m}^{\omega,\omega'}(a,b;c| c,d;e)= } \nonumber \\ 
& &  \displaystyle{\sum_{n=\sup(0,[\omega'])}^{n=\inf(m,[\omega])}} x_{m}^{\omega}(\inf(a,b),\inf(\sup(a,b),c))
x_{m-n}^{\omega'}(\inf(c,d),\inf(\sup(c,d),e)), \nonumber \\
\end{eqnarray}
with
\begin{equation}   \label{eq:OmegfOmegaprmf}
\omega=a+b+c-2\sup(a,b,c) \mbox{\hspace{0.5em}and\hspace{0.5em}} \omega'=c+d+e-2\sup(c,d,e).
\end{equation}
Thus, degree $\Omega$ is given by:
\begin{equation}   \label{eq:DegreeOmega}
\Omega = [a+b+c]+[c+d+e]-2\sup(a,b,c)-2\sup(c,d,e).
\end{equation} 
Product ${\cal P}^{\,\omega_{f}}\bigl(\inf(b,d),\inf(\sup(b,d),f)\bigr) {\cal P}^{\,\omega'_{f}}\bigl(\inf(a,f),\inf(\sup(a,f),e)\bigr)$ may be expressed by means of formulas similar to 
(\ref{eq:ProdPabcPcde})-(\ref{eq:DEfCoeffy}), with $\Omega_ {f}=[\omega_f]+[\omega'_{f}]$, and
\begin{equation}   \label{eq:OmegaOmegaprm}
\omega_{f}=b+d+f-2\sup(b,d,f) \mbox{\hspace{0.5em}and\hspace{0.5em}} \omega'_{f}=a+e+f-2\sup(a,e,f).
\end{equation}
Thus, for any $f$, degree $\Omega_ {f}$ is given by:
\begin{equation}   \label{eq:DegreeOmegaf}
\Omega_{f} = [b+d+f]+[a+e+f]-2\sup(b,d,f)-2\sup(a,e,f).
\end{equation} 

   \subsection{Special case with $d_{0}$ as frontal factor}
Looking at eq.~(\ref{eq:PreliminDeltaSSumRule}), we first must check its coherence when $d_{0}$ factorizes the left-hand side. This case arises only if the ``perimeters" ($a+b+c$) and ($c+d+e$) have different $\tau$-parities, {\em i.e.} one perimeter integral and the other one half-integral, which
leads to
\begin{equation}   \label{eq:Paritytriangleleft}
\tau_{a+b+c}+\tau_{c+d+e}=0+1=1+0=1.
\end{equation}
This implies that $b+d$ and $a+e$ have also different $\tau$-parities. Therefore, $\forall f$, integral or half-integral, on the right-hand side of 
eq.~(\ref{eq:PreliminDeltaSSumRule}), we have also
\begin{equation}   \label{eq:Paritytrianglerightf}
\tau_{b+d+f}+\tau_{a+e+f}=1.
\end{equation}
Consequently $d_{0}$ {\sl can be canceled on both sides}. Then eq.~(\ref{eq:PreliminDeltaSSumRule}) here becomes
\begin{eqnarray} \label{eq:DeltaSSumRulewithd0}
(-1)^{2\sup(a,b,c)\tau_{a+b+c} + 2\sup(c,d,e)\tau_{c+d+e} }\begin{array}{c} \frac {(2\sup(a,b,c))!(2\sup(c,d,e))!} { (2c)!} \end{array}
\nonumber \\
\times  ( \gamma_{\,\sup(a,b,c)} \cdots \gamma_{c+\frac{1}{2}} )  
( \gamma_{\,\sup(c,d,e)} \cdots \gamma_{e+\frac{1}{2}} ) {\cal R}^{\Omega}(a,b;c|c,d;e)
\bigtriangleup^{S}(a\, b \,c)\bigtriangleup^{S}(c\, d \,e)  \nonumber \\
= (-1)^{[a+b+d+e]}  \sum_{f} (-1)^{2\sup(b,d,f)\tau_{b+d+f} + 2\sup(a,f,e)\tau_{a+f+e}} 
\left\{ \begin{array}{ccc} a & b & c  \\ d & e & f \end{array} \right\}^{\!_S} \nonumber \\
\times \begin{array}{c} \frac {(2\sup(b,d,f))!(2\sup(a,f,e))!} { (2f)!} \end{array} 
( \gamma_{\,\sup(b,d,f)} \cdots \gamma_{f+\frac{1}{2}} )
( \gamma_{\,\sup(a,f,e)} \cdots \gamma_{e+\frac{1}{2}} ) \nonumber \\
\times  \bigtriangleup^{S}(b\, d \,f)\bigtriangleup^{S}(a\, f \,e) {\cal R}^{\Omega_{f}}(b,d;f|a,f;e). \nonumber \\ 
\end{eqnarray}
In the present case, the degrees $\Omega$ and $\Omega_{f}$ are the following:
\begin{eqnarray}   \label{eq:DegOmegaandOmegaf}
\begin {array}{c}
\Omega = [a+b+d+e]+2c -2\sup(a,b,c) - 2\sup(c,d,e) \\
\Omega_{f} = [a+b+d+e]+2f-2\sup(b,d,f) - 2\sup(a,f,e) \end{array}.
\end{eqnarray}

   \subsection{The other cases} They are in number of two for the left-hand side of eq.~(\ref{eq:PreliminDeltaSSumRule}).
Either $a+b+c$ and $c+d+e$ are integral, or $a+b+c$ and $c+d+e$ are half-integral. This implies that $a+b$ and $d+e$ have the same $\tau$-parity,
{\em i.e.} integral or half-integral.\par
For convenience, let us now define a general phase factor $\varphi$ as follows:
\begin{eqnarray}   \label{eq:Defphasephi}
(-1)^{\varphi (a+b+d+e;c)}=(-1)^{[a+b+d+e]-\tau_{a+b+c}\tau_{c+d+e}}  \nonumber \\ 
= (-1)^{[a+b+d+e]+\tau_{c}(1-\tau_{a+b+d+e})}.
\end{eqnarray} 
                                In these cases eq.~(\ref{eq:PreliminDeltaSSumRule}) takes the form:
\begin{eqnarray} \label{eq:DeltaSSumRulewithnod0}
\begin{array}{c} \frac {(2\sup(a,b,c))!(2\sup(c,d,e))!} { (2c)!} \end{array} 
( \gamma_{\,\sup(a,b,c)} \cdots \gamma_{c+\frac{1}{2}} )   
( \gamma_{\,\sup(c,d,e)} \cdots \gamma_{e+\frac{1}{2}} ) \nonumber \\
\times  
\bigtriangleup^{S}(a\, b \,c)\bigtriangleup^{S}(c\, d \,e)  \nonumber \\
\times 
\left\{ \begin{array}{l}
\begin{array}{lcr} 
& {\cal R}^{\Omega}(a,b;c|c,d;e) & 
\begin{array}{c} \mbox{\hspace{0.5em} {\scriptsize if  $c$ and $(a+b,d+e)$ \hspace{0.3em} have identical $\tau$-parities}}  \\
\mbox{\scriptsize case where \hspace{0.3em}} \scriptstyle (a+b+c,c+d+e) \mbox{\hspace{0.3em} {\scriptsize integral} }  \end{array} \\
\\ [0.1em]
{d_{0}}^{2} & {\cal R}^{\Omega}(a,b;c|c,d;e) & 
\begin{array}{c} \mbox{\hspace{0.5em} {\scriptsize if $c$ and $(a+b,d+e)$ \hspace{0.3em} have different $\tau$-parities }} \\
\mbox{\scriptsize case where \hspace{0.3em}} \scriptstyle (a+b+c,c+d+e) \mbox{\hspace{0.3em} {\scriptsize half-integral} }  
\end{array}  
\end{array}  
\end{array} \right. \nonumber \\
= (-1)^{\varphi (a+b+d+e;c)}  \sum_{f} (-1)^{2\sup(b,d,f)\tau_{b+d+f} + 2\sup(a,f,e)\tau_{a+f+e}} 
\left\{ \begin{array}{ccc} a & b & c  \\ d & e & f \end{array} \right\}^{\!_S} \nonumber \\
\times \begin{array}{c} \frac {(2\sup(b,d,f))!(2\sup(a,f,e))!} { (2f)!} \end{array} 
( \gamma_{\,\sup(b,d,f)} \cdots \gamma_{f+\frac{1}{2}} ) 
( \gamma_{\,\sup(a,f,e)} \cdots \gamma_{e+\frac{1}{2}} ) \nonumber \\
\times  \bigtriangleup^{S}(b\, d \,f)\bigtriangleup^{S}(a\, f \,e) \nonumber \\ 
\times 
\left\{ \begin{array}{l}
\begin{array}{lcr} 
& {\cal R}^{\Omega_{f}}(b,d;f|a,f;e) & 
\begin{array}{c} \mbox{\hspace{0.5em} {\scriptsize if  $f$ and $(b+d,a+e)$ \hspace{0.3em} have identical $\tau$-parities }} \\ 
\mbox{\scriptsize case where \hspace{0.3em}} \scriptstyle (b+d+f,a+e+f) \mbox{\hspace{0.3em} {\scriptsize integral} }  \end{array}
\\ [1em] 
{d_{0}}^{2} & {\cal R}^{\Omega_{f}}(b,d;f|a,f;e) & 
\begin{array}{c} \mbox{\hspace{0.5em} {\scriptsize if $f$ and $(b+d,a+e)$ \hspace{0.3em} have different $\tau$-parities }} \\
\mbox{\scriptsize case where \hspace{0.3em}} \scriptstyle (b+d+f,a+e+f) \mbox{\hspace{0.3em} {\scriptsize half-integral} }  
\end{array}
\end{array} 
\end{array} \right. \nonumber \\
\end{eqnarray} 
The relevant degrees are the following:
\begin{equation}   \label{eq:Degreeabccdeintegr}
\mbox{\hspace*{1em}}\Omega = [a+b+d+e]+2c - 2\sup(a,b,c) - 2\sup(c,d,e)- \tau_{c}(1-\tau_{a+b+d+e}),
\end{equation}
and 
\begin{equation} \label{eq:DegreeOmegasubcases}
\mbox{\hspace*{1em}}\Omega_{f}=[a+b+d+e]+2f -2\sup(b,d,f)-2\sup(a,f,e)- \tau_{f}(1-\tau_{a+b+d+e}).
\end{equation}
\underline{Remarks}: Note that, for $a,b,d,e$ given, the summation over $f$ by step $\frac{1}{2}$ changes the $\tau$-parity of $f$, which becomes
integral or half-integral, half the time {\em etc}, then factor ${d_{0}}^{2}$ appears, or doesn't appear as frontal factor of polynomials
${\cal R}$, whose the degrees $\Omega_{f}$ simultaneously follow the associated variations.  \par
\underline{Special notations and definitions}: 
In order to continue the mathematical analysis in progress, we need to define some simple notations which
specify {\em explicitly} if the step of running indices involved in summations is $\frac{1}{2}$, and if
these indices can be integral {\em and} half-integral, or only integral or only half-integral. Namely:\par
$\sumhalf{f}{}$ stands for: step=$\frac{1}{2}$, and (of course here) $f$ integral and half-integral.\par
$\displaystyle \sum_{\overline{f}}$ stands for: step=1 and $\overline{f}$ integral, 
\hspace{1em} $\displaystyle \sum_{\underline{f}}$ stands for: step=1 and $\underline{f}$ half-integral.\par
In other words, let $\{f\}$ be the set of the values taken by $f$, then it can be partitioned in the following direct sum:
\begin{equation}   \label{eq:Partitofindex}
\{f\} = \{\overline{f}\} \oplus \{\underline{f}\}. 
\end{equation}
Thus a standard summation for osp(1$|$2) can be separated as follows:
\begin{equation}   \label{eq:Decomposp12Sums}
\sumhalf{f}{} = \sum_{\overline{f}} + \sum_{\underline{f}}. 
\end{equation}
Otherwise no specification is necessary for customary formulas regarding su(2).

                              \subsection{Form of polynomials in $(c_{0}+{d_{0}}^{2}), {d_{0}}^{2}$ coming from the $\gamma$'s products}
Let us define a polynomial homogeneous in variables $(c_{0}+{d_{0}}^{2}),{d_{0}}^{2}$, of degree $\Omega_{\Gamma}$, with coefficients depending on two parameters $\kappa_{>}$ and $\kappa_{<}$, satisfying  $\kappa_{<} \leq \kappa_{>}$, as follows:
\begin{equation}   \label{eq:PolynZfromgamas}
{\cal Z}^{\Omega_{\Gamma}}(\kappa_{>};\kappa_{<})=
\prod_{l=[\kappa_{<}]}^{l=[\kappa_{>}-\frac{1}{2}]} \left( (2l+1)^{2}(c_{0}+{d_{0}}^{2})-{d_{0}}^{2}\right).
\end{equation}
According to an obvious {\em convention}, the finite product $\prod$ equals $1$ and ${\Omega_{\Gamma}}=0$ if
$[\kappa_{<}]> [\kappa_{>}-\frac{1}{2}]$, for instance if $[\kappa_{<}]= [\kappa_{>}-\frac{1}{2}]+1$, this case arising only 
if $\kappa_{<}=\kappa_{>}$ are integral. If $l$ reduces to the single value $l=0$, case where
$[\kappa_{<}]=[\kappa_{>}-\frac{1}{2}]=0$, then $\Omega_{\Gamma}=1$ and ${\cal Z}^{1}(\kappa_{>};\kappa_{<})=c_0=\gamma_{\frac{1}{2}}$.   
Thus polynomials ${\cal Z}$ have the following degree:
\begin{equation}   \label{eq:ValueofOmegakappa}
\Omega_{\Gamma}=\begin{array}{c}[\kappa_{>}+\frac{1}{2}]\end{array}-[\kappa_{<}].
\end{equation}
Thanks to these definitions, it can checked that the product $(\gamma_{\kappa_{>} }\cdots \gamma_{\kappa_{<} })$ may be represented by an
analytical expression involving double factorials such as:
\begin{eqnarray}   \label{eq:ProductsofGammas}
\lefteqn{ 
( \gamma_{\kappa_{>} }\cdots \gamma_{\kappa_{<}} )=  } \nonumber \\
& & \frac{(2[\kappa_{>}])!!(2[\kappa_{<}]-1)!!}{(2[\kappa_{<}-\frac{1}{2}])!!(2[\kappa_{>}+\frac{1}{2}]-1)!!} 
\; (c_{0}+{d_{0}}^{2})^{[\kappa_{>}]-[\kappa_{<}-\frac{1}{2}]}\; {\cal Z}^{\Omega_{\Gamma}}(\kappa_{>};\kappa_{<}). \nonumber \\
\end{eqnarray}
\hspace{1.5em}This important section (with Appendix B as its very valuable companion) is the last of our paper. That sets precisely the problem to solve, thanks to a lot of definitions, notations, equations or tools necessary for an identification process concerning various polynomials, {\em unknown} in scientific literature. However the complexity of the task remaining to  achieve, as well as the paradoxical ``warning" stressed at the begining of the section, force us to close here our analysis and to present our work as an ``open problem".

                                      \section{Conclusion}\label{Conclus}

\hspace*{1.5em}After noting that a theoretical attempt was made twenty years ago by Zeng \cite{Zeng.4}, which suggested a kind of osp(1$|$2)-sum rule by means of the $\bigtriangledown$-triangles of su(2), the developments found in the present paper let appear that we do not know how his proposition could be assessed, see our analysis in Appendix~C.\par
\hspace*{1.5em}Let us now summarize the main features of our work.\par
Analytical expressions for the cases su(2) and su$_q$(2) have been obtained with easiness. Regarding osp(1$|$2), 
the ultimate identification process over two independent parameters, namely $c_0+{d_0}^2$ and  ${d_0}^2$, also should lead to
a specific triangle sum rule. Unfortunately, it seems to come out onto an analytical dead end.\par
The best should be to try doing hand-calculations for low values of spins occuring in $6$-$j^{S}$ symbols (up to 20 for example). However, that involves the availability of $6$-$j^{S}$ tables, listed in terms of prime numbers, as usual for classical
su(2) $6$-$j$ symbols. {\em These tables have neither been programmed nor published up today}. For any interested and motivated searcher / programmer, this task could be rapidly carried out from our general formulas given in 
ref.~\cite{L.B.Nuovo.I}. That could furnish an invaluable help not only for the problem under study but certainly also in other areas of the physics.\par
At least, that could allow ones to anticipate what kind of identities can occur between expansion coefficients $x$ whose analytical expression seems so difficult to obtain.\par
May be, that can lead to a failure. Taken into account that all our equations are rigorous, one can wonder if it is really ``licit"      
to do identifications over tensor operators like $\mbox{\boldmath$S$}^{c}$. Indeed there are a few number of them which are
``zero" by construction, as proved by two theorems inside this paper. Certainly this is not the most important aspect.\par
On the other hand, we have restricted to $c_0$ and $d_0$ to be real numbers in our study. May be an extension to imaginary numbers must be taken into account?\par
Whatsoever the final solution, the yet simple hypothesis of a triangle sum rule for 
osp(1$|$2) involves a lot of problems, including rigorous proof of weak conjectures, like the fact that $x$ coefficients are integral numbers (see sect.~\ref{ClosureSetA}) and functional relations between polynomials 
${\cal P}^{\,\omega}$ and  ${\cal Q}^{\,\omega}$ (see sect.~\ref{ClosureSetB}). That displays a rather vast and promising work remaining to achieve.         

\appendix

                               \renewcommand{\theequation}{A.\arabic{equation}}
                               \setcounter{equation}{0}
 
                   \section*{Appendix A \\ [1em] Study on $\mbox{\boldmath$S$}^{\frac{1}{2}}$ and $\mbox{\boldmath$S$}^{1}$: reduced matrix elements and some properties}\label{RedMatrhalfone}
                                 
\hspace*{1.5em}This study is interesting because it is found that coefficients $\alpha$ and $\gamma$ introduced by necessity (see sect.~\ref{TensorIter})
for our tensorial iterative process (see sect.~\ref{FirstTensFormulasH}) occur also in another field. They appear simply and straightforwardly when looking at the primitive
spinor $\mbox{\boldmath$S$}^{\frac{1}{2}}$, and $\mbox{\boldmath$S$}^{1}$, which is an alternative generator of osp(1$|$2) (under conditions
related to $c_{0}+{d_{0}}^{2}$).\par
General analytical formulas for any $(j\|\;\|j')$ reduced matrix elements 
of tensor products may be found clearly explicited in our recent paper \cite{L.B.Nuovo.I}. Spin $j$ refers to a standard 
representation basis $|j m>$ issued from the super-angular momentum ${\cal J}^{1}$. Tensor operators such as 
$\mbox{\boldmath$S$}^{\frac{1}{2}}$ or $\mbox{\boldmath$S$}^{1}$ of course are irreducible with respect to ${\cal J}^{1}$. 
\par
First we have:
\begin{equation}   \label{eq:RedMatrSzero}
(j\|\mbox{\boldmath$S$}^{0}\|j')=\delta_{j,j'}.
\end{equation}

                               \renewcommand{\thesubsection}{A.\arabic{subsection}}

     \subsection{Reduced matrix elements of  $\mbox{\boldmath$S$}^{\frac{1}{2}}$}\label{RdMatrS12}
Evaluation of the reduced matrix elements of  the product
$[\mbox{\boldmath$S$}^{\frac{1}{2}}\times\mbox{\boldmath$S$}^{\frac{1}{2}}]^{0}=c_{0}\mbox{\boldmath$S$}^{0}$ yields
\begin{eqnarray}   \label{eq:firstRelforSzero}
\lefteqn{ c_{0}= } \nonumber \\
&  \begin{array}{c}(j\|\mbox{\boldmath$S$}^{\frac{1}{2}}\|j\mbox{$-$}\frac{1}{2})(j\mbox{$-$}\frac{1}{2})\|\mbox{\boldmath$S$}^{\frac{1}{2}}\|j)
\mbox{$-$}{(j\|\mbox{\boldmath$S$}^{\frac{1}{2}}\|j)}^{2}
\mbox{$-$}(j\|\mbox{\boldmath$S$}^{\frac{1}{2}}\|j\mbox{$+$}\frac{1}{2})(j\mbox{$+$}\frac{1}{2})\|\mbox{\boldmath$S$}^{\frac{1}{2}}\|j).  \end{array}
\end{eqnarray}
After that, the evaluation of the reduced matrix elements of         
$[\mbox{\boldmath$S$}^{\frac{1}{2}}\times\mbox{\boldmath$S$}^{\frac{1}{2}}]^{\frac{1}{2}}=d_{0}\sqrt{2}\mbox{\boldmath$S$}^{\frac{1}{2}}$
furnishes only two relevant equations:
\begin{eqnarray}   \label{eq:firstRelforSonehalf}
\lefteqn{ 2d_{0}= } \nonumber \\
& & 
\begin{array}{c}\frac{(-1)^{2j} }{2j} \left( \sqrt{2j(2j-1)}
(j\mbox{$-$}\frac{1}{2}\|\mbox{\boldmath$S$}^{\frac{1}{2}}\|j\mbox{$-$}\frac{1}{2})
\mbox{$-$}\sqrt{(2j+1)2j}
(j\|\mbox{\boldmath$S$}^{\frac{1}{2}}\|j) \right),
\end{array} 
\end{eqnarray}
and
\begin{eqnarray}   \label{eq:NondiagRedMatrS12}
\lefteqn{ 2d_{0}= (-1)^{2j}\frac{1}{\sqrt{2j(2j+1)}} } \nonumber \\   
&& \times \left\{ \begin{array}{llr} \label{eq:firstRelforSonehalfjj}
& & 
(2j \mbox{$+$} 1)  
(j\|\mbox{\boldmath$S$}^{\frac{1}{2}}\|j\mbox{$-$}\frac{1}{2})(j\mbox{$-$}\frac{1}{2}\|\mbox{\boldmath$S$}^{\frac{1}{2}}\|j) \\
 & & \mbox{$-$} (4j \mbox{$+$} 1) {(j\|\mbox{\boldmath$S$}^{\frac{1}{2}}\|j)}^{2} \\
& & \mbox{$-$} 2j
(j\|\mbox{\boldmath$S$}^{\frac{1}{2}}\|j\mbox{$+$}\frac{1}{2})(j\mbox{$+$}\frac{1}{2}\|\mbox{\boldmath$S$}^{\frac{1}{2}}\|j).
\end{array} \right. \nonumber \\
\end{eqnarray}
\hspace*{1.5em}Solution of eq.~(\ref{eq:firstRelforSonehalf}) is obtained by following a recursive method, thus we find
\begin{equation}   \label{eq:SolEqB1}
(j\|\mbox{\boldmath$S$}^{\frac{1}{2}}\|j)=(-1)^{2j+1}\frac{(2j+\tau_{j})}{\sqrt{2j(2j+1)}} d_{0},
\end{equation}
which is just our $\alpha$ coefficient introduced previously in eq.~(\ref{eq:DefCoeffalpha}). Consequently we have
\begin{equation}   \label{eq:RedMatSonehalfjj}
(j\|\mbox{\boldmath$S$}^{\frac{1}{2}}\|j)=\alpha_{j}.
\end{equation}
\hspace*{1.5em}Solution of eq.~(\ref{eq:NondiagRedMatrS12}) may be derived from the preceding result and eq.~(\ref{eq:firstRelforSzero}):
\begin{equation}   \label{eq:SolfornondiagRedMatrS12}
\begin{array}{c}(j\|\mbox{\boldmath$S$}^{\frac{1}{2}}\|j\mbox{$-$}\frac{1}{2})(j\mbox{$-$}\frac{1}{2}\|\mbox{\boldmath$S$}^{\frac{1}{2}}\|j)\end{array}
= -
\left\{ \begin{array}{lr}
2j(c_{0}+{d_{0}}^{2}) & \mbox{\scriptsize (j integral)}\\
\displaystyle 2j(c_{0}+{d_{0}}^{2})-\frac{{d_{0}}^{2}}{2j} & \mbox{\scriptsize (j half-integral)} \end{array} \right. .
\end{equation}
We recognize our $\gamma$ coefficient introduced previously in eq.~(\ref{eq:DefCoeffgamma}). Therefore we have
\begin{equation}   \label{eq:SolEqB2}
\begin{array}{c}(j\|\mbox{\boldmath$S$}^{\frac{1}{2}}\|j\mbox{$-$}\frac{1}{2})(j\mbox{$-$}\frac{1}{2}\|\mbox{\boldmath$S$}^{\frac{1}{2}}\|j)\end{array}
= - \gamma_{j}.
\end{equation}
It can be noticed also that eq.~(\ref{eq:firstRelforSzero}) now may be rewritten as a recursion relation for $\gamma$:
\begin{equation}   \label{eq:Recursforgamma}
c_{0}+{\alpha_{j}}^{2}=\gamma_{j+\frac{1}{2}} - \gamma_{j}.
\end{equation}
					 \renewcommand{\thesubsection}{A.\arabic{subsection}}
					
     \subsection{Reduced matrix elements of  $\mbox{\boldmath$S$}^{1}$}\label{RdMatrS1}
Calculations necessary for $\mbox{\boldmath$S$}^{1}$ don't offer particular difficulties. Let us list below all relevant results. They are derived from the defining expression $\mbox{\boldmath$S$}^{1}=[\mbox{\boldmath$S$}^{\frac{1}{2}}\times\mbox{\boldmath$S$}^{\frac{1}{2}}]^{1}$. 
\begin{equation}   \label{eq:RMS1jj-1}
(j\|\mbox{\boldmath$S$}^{1}\|j\mbox{$-$}1)=
\begin{array}{c}(j\|\mbox{\boldmath$S$}^{\frac{1}{2}}\|j\mbox{$-$}\frac{1}{2})
(j\mbox{$-$}\frac{1}{2}\|\mbox{\boldmath$S$}^{\frac{1}{2}}\|j\mbox{$-$}1),
\end{array}
\end{equation}
\begin{equation}   \label{eq:RMS1j-1j}
(j\mbox{$-$}1\|\mbox{\boldmath$S$}^{1}\|j)=
\begin{array}{c}(j\mbox{$-$}1\|\mbox{\boldmath$S$}^{\frac{1}{2}}\|j\mbox{$-$}\frac{1}{2})
(j\mbox{$-$}\frac{1}{2}\|\mbox{\boldmath$S$}^{\frac{1}{2}}\|j).
\end{array}
\end{equation}
Both next formulas are interesting because they involve the factor $\tau$ previously defined by eq.~(\ref{eq:taudef}):
\begin{equation}   \label{eq:RMS1jj-12}
\begin{array}{c}(j\|\mbox{\boldmath$S$}^{1}\|j\mbox{$-$}\frac{1}{2})\end{array}= -
\begin{array}{c}(j\|\mbox{\boldmath$S$}^{\frac{1}{2}}\|j\mbox{$-$}\frac{1}{2})\end{array}
\frac{d_{0}\sqrt{2}}{\sqrt{(2j-1)(2j+1)}} \tau_{j-\frac{1}{2}},
\end{equation}
\begin{equation}   \label{eq:RMS1j-12j}
\begin{array}{c}(j\mbox{$-$}\frac{1}{2}\|\mbox{\boldmath$S$}^{1}\|j)\end{array}= -
\begin{array}{c}(j\mbox{$-$}\frac{1}{2}\|\mbox{\boldmath$S$}^{\frac{1}{2}}\|j)\end{array}
\frac{d_{0}\sqrt{2}}{\sqrt{(2j-1)(2j+1)}} \tau_{j-\frac{1}{2}}.
\end{equation}
Factor $\tau$ occurs in the calculation when we need to evaluate the following expression involving two different $\alpha$'s:
\begin{equation}   \label{eq:TauOccur}
\alpha_{j-\frac{1}{2}}\sqrt{\frac{2j+1}{2j}}+\alpha_{j}\sqrt{\frac{2j-1}{2j}} = \frac{2d_{0}}{\sqrt{(2j-1)(2j+1)}} \tau_{j-\frac{1}{2}}.
\end{equation} 
Eqs.~(\ref{eq:RMS1jj-12})-(\ref{eq:RMS1j-12j}) imply a rather unexpected result, namely:
\begin{equation}   \label{eq:ZeroRedMatrS1}
\begin{array}{c}(j\|\mbox{\boldmath$S$}^{1}\|j\mbox{$-$}\frac{1}{2})\end{array}=
\begin{array}{c}(j\mbox{$-$}\frac{1}{2}\|\mbox{\boldmath$S$}^{1}\|j)\end{array}=0,
\mbox{\hspace{2em}}\forall \, j \,\mbox{half-integral}.
\end{equation}
All calculations done, the last relevant reduced matrix element has the form:
\begin{equation}   \label{eq:RMS1jj}
(j\|\mbox{\boldmath$S$}^{1}\|j)=-\sqrt{2}\sqrt{2j(2j+1)}\left(c_{0}+{d_{0}}^{2}\right).
\end{equation}
Here again this results from an expression involving $\alpha$ and $\gamma$ whose computation was necessary, namely:
\begin{equation}   \label{eq:Idalphagamma}
(2j+1)\gamma_{j}+{\alpha_{j}}^{2}+2j\gamma_{j+\frac{1}{2}}=4j(2j+1)\left(c_{0}+{d_{0}}^{2}\right).
\end{equation}

      \subsection{Some remarks about the phases}\label{PhaseRemarks}
As it has been seen, the cases $(j\|\mbox{\boldmath$S$}^{\frac{1}{2}}\|j)$ and $(j\|\mbox{\boldmath$S$}^{1}\|j)$ being considered as special, none of all the other reduced matrix elements, non $j$-diagonal, like
$(j\|\mbox{\boldmath$S$}^{\frac{1}{2}}\|j')$ or $(j\|\mbox{\boldmath$S$}^{1}\|j')$, with $j \neq j'$, has been written as a general explicit function
of $\gamma$. Indeed, even if we assume reasonably that
\begin{equation}   \label{eq:HypothPhasejjprime}
|(j\|\mbox{\boldmath$S$}^{\frac{1}{2}}\|j')|=|(j'\|\mbox{\boldmath$S$}^{\frac{1}{2}}\|j)|,
\end{equation}
and examine the case where $(j\|\mbox{\boldmath$S$}^{\frac{1}{2}}\|j')$ is a {\em real} number, a problem of phases is encountered, which depends on the relative positions of $c_{0}$ and ${d_{0}}^{2}$. Two classes may be defined, with a binary variable $\epsilon$ (with value $0$ or $1$) as follows: 
\begin{equation}   \label{eq:Classr}
\begin{array}{c}(j\|\mbox{\boldmath$S$}^{\frac{1}{2}}\|j\mbox{$-$}\frac{1}{2})\end{array}
=(-1)^{\epsilon_{r}}\begin{array}{c}|(j\|\mbox{\boldmath$S$}^{\frac{1}{2}}\|j\mbox{$-$}\frac{1}{2})|\end{array},
\end{equation}   
\begin{equation}   \label{eq:Classl}
\begin{array}{c}(j\mbox{$-$}\frac{1}{2}\|\mbox{\boldmath$S$}^{\frac{1}{2}}\|j)\end{array}
=(-1)^{\epsilon_{l}}\begin{array}{c}|(j\mbox{$-$}\frac{1}{2}\|\mbox{\boldmath$S$}^{\frac{1}{2}}\|j)|\end{array}.
\end{equation}
Let us proceed to briefly analyze two elementary examples. First, suppose that $c_{0}>0$. Then $\gamma_{j}>0 \; \forall\, j$ and
from eq.~(\ref{eq:SolEqB2}) we derive $(-1)^{\epsilon_{r}+\epsilon_{l}}=-1$. Thus, in this case
\begin{equation}   \label{eq:FirstExmpl}
\begin{array}{c}(j\|\mbox{\boldmath$S$}^{\frac{1}{2}}\|j\mbox{$-$}\frac{1}{2})\end{array}=\pm \sqrt{\gamma_{j}}
\mbox{\hspace{0.5em} and \hspace{0.5em}}\begin{array}{c}(j\mbox{$-$}\frac{1}{2}\|\mbox{\boldmath$S$}^{\frac{1}{2}}\|j)\end{array}=
\mp \sqrt{\gamma_{j}}.
\end{equation}
Secondly, if $c_{0}+{d_{0}}^{2}<0$, then $\gamma_{j}<0 \; \forall\, j$, $(-1)^{\epsilon_{r}+\epsilon_{l}}=+1$, whence      
\begin{equation}   \label{eq:SecondExmpl}
\begin{array}{c}(j\|\mbox{\boldmath$S$}^{\frac{1}{2}}\|j\mbox{$-$}\frac{1}{2})\end{array}=\pm \sqrt{-\gamma_{j}}
\mbox{\hspace{0.5em} and \hspace{0.5em}}\begin{array}{c}(j\mbox{$-$}\frac{1}{2}\|\mbox{\boldmath$S$}^{\frac{1}{2}}\|j)\end{array}=
\pm \sqrt{-\gamma_{j}}.
\end{equation}
There are other examples not so simple, but out of the scope of this paper.

 \subsection{$\mbox{\boldmath$S$}^{1}$ as another generator for osp(1$|$2)}\label{S1Generator}
The standard generator of osp(1$|$2), {\em i.e.} the super-angular momentum $\mbox{\boldmath$\cal J$}^{1}$ itself , satisfies intrinsic commutation relations \cite{L.B.Nuovo.I}:
\begin{equation}   \label{eq:IntrinsIrrJ1}
[\mbox{\boldmath$\cal J$}^{1}\times \mbox{\boldmath$\cal J$}^{1}]^{1}=-\frac{1}{2}\sqrt{\frac{3}{2}}\mbox{\boldmath$\cal J$}^{1}.
\end{equation}
From eq.~(\ref{eq:Casepi1}) with $\lambda=1$ and $\kappa=1$ we have
\begin{equation}   \label{eq:IntrinsIrrS1}
[\mbox{\boldmath$S$}^{1}\times \mbox{\boldmath$S$}^{1}]^{1}=2\sqrt{3}\left( c_{0}+{d_{0}}^{2} \right) \mbox{\boldmath$S$}^{1}.
\end{equation}
Then, in order to be a generator of osp(1$|$2), $\mbox{\boldmath$S$}^{1}$, actually depending on $c_{0},\, d_{0}$, must satisfy conditions such as given by
eq.~(\ref{eq:IntrinsIrrJ1}), which leads to the following equation:
\begin{equation}   \label{eq:S1becomesGenerator}
\left( c_{0}+{d_{0}}^{2} \right)=-\frac{1}{4} \sqrt{\frac{1}{2}}.
\end{equation}
In conclusion of the present topic, 
${\mbox{\boldmath$S$}^{1}}
_{|_{c_{0}+{d_{0}}^{2}=-\frac{1}{4} \sqrt{\frac{1}{2}}}}$ is another valid generator for osp(1$|$2).

                               \renewcommand{\theequation}{B.\arabic{equation}}
                               \setcounter{equation}{0}

                 \section*{Appendix B \\ [1em] Polynomials ${\cal P}^{\,\omega}$ and  ${\cal Q}^{\,\omega}$: some properties and recursion relations} \label{PQpolynoms}

Notations used: $p$ for indicating that $\omega=p$ is integral, $\omega=\pi$ half-integral. \par
                                \setcounter{subsection}{0}      
				\renewcommand{\thesubsection}{B.\arabic{subsection}}
				\subsection{Polynomials ${\cal P}^{\,\omega}$ and general relations}

The detailed forms of eq.~(\ref{eq:ReccurSHigh}) are found to be:
\begin{eqnarray}   \label{eq:RecurrPpIter} 
\lefteqn{
\frac{(2\kappa-p)!}{(2\kappa)!} {\cal P}^{\,p}(\lambda,\kappa)= }  \nonumber \\
& &
\begin{array}{c}
{\displaystyle \frac{(2\kappa-p)!}{(2\kappa)!}}
\left(\displaystyle \frac{(-1)^{2\kappa+1} \alpha_{\kappa}}{ \sqrt{2\kappa(2\kappa+1)}} \right) 
{\cal P}^{\,p-\frac{1}{2}}(\lambda- \frac{1}{2},\kappa) \\ [1em]
+(2\lambda+2\kappa-p) {\displaystyle \frac{(2\kappa-p)!}{(2\kappa)!}} \gamma_{\kappa}
{\cal P}^{\,p-1}(\lambda- \frac{1}{2},\kappa- \frac{1}{2}) \\ [1em]
+{\displaystyle \frac{(2\kappa+1-p)!}{(2\kappa+1)!}}
{\cal P}^{\,p}(\lambda- \frac{1}{2},\kappa+ \frac{1}{2}),
\end{array}\nonumber \\ 
\end{eqnarray}
and
\begin{eqnarray}   \label{eq:RecurrPpiIter}
\lefteqn{
\frac{(2\kappa-\pi-\frac{1}{2})!}{(2\kappa)!} {\cal P}^{\,\pi}(\lambda,\kappa)= }  \nonumber \\
& &
\begin{array}{c}
(2\lambda+2\kappa-\pi+\frac{1}{2})
{\displaystyle \frac{(2\kappa-\pi+\frac{1}{2})!}{(2\kappa)!}}
\left(\displaystyle \frac{(-1)^{2\kappa+1} \alpha_{\kappa}}{ \sqrt{2\kappa(2\kappa+1)}} \right) 
{\cal P}^{\,\pi-\frac{1}{2}}(\lambda- \frac{1}{2},\kappa) \\ [1em]
- (2\lambda+2\kappa-\pi+\frac{1}{2}){\displaystyle \frac{(2\kappa-\pi-\frac{1}{2})!}{(2\kappa)!}} \gamma_{\kappa}
{\cal P}^{\,\pi-1}(\lambda- \frac{1}{2},\kappa- \frac{1}{2}) \\ [1em]
-{\displaystyle \frac{(2\kappa-\pi+\frac{1}{2})!}{(2\kappa+1)!}}
{\cal P}^{\,\pi}(\lambda- \frac{1}{2},\kappa+ \frac{1}{2}).
\end{array} \nonumber \\ 
\end{eqnarray}
These formulas, such as they stand, are sufficient for asserting that any polynomial ${\cal P}^{\,\omega}$ (where $\omega$ is integral or half-integral) can be written under the most general form as:
\begin{equation}   \label{eq:FormofPolynP}
{\cal P}^{\,\omega}(\lambda,\kappa)=\left( {d_{0}} \right)^{\tau_{\omega}}
\sum_{m=0}^{m=[\omega]} x_{\,m}^{\,\omega}(\lambda,\kappa)
\left(c_{0}+{d_{0}}^{2}\right)^{[\omega]-m}({d_{0}}^{2})^{m}.
\end{equation}
This means that ${\cal P}^{\,\omega}(\lambda,\kappa)$ is the product of $\left( {d_{0}} \right)^{\tau_{\omega}}$ 
by {\em a polynomial of degree $[\omega]$, homogeneous in $\left(c_{0}+{d_{0}}^{2}, {d_{0}}^{2}\right)$ }.
$[\omega]$  denotes the integral part of $\omega$.\par
For instance, from sect.~\ref{FirstTensFormulasH}, the first $x_{\,m}^{\,\omega}(\lambda,\kappa)$'s are given by: 
\begin{equation}   \label{eq:Firstxomega}
\begin{array}{l}
{\scriptstyle x_{\,0}^{\,0}(\lambda,\kappa)=1,} \\ [0.5em]
{\scriptstyle x_{\,0}^{\,\frac{1}{2}}(\lambda,\kappa)= 2\lambda\tau_{\kappa}+2\kappa\tau_{\lambda},} \\ [0.5em]
{\scriptstyle x_{\,0}^{\,1}(\lambda,\kappa)=2\lambda\cdot 2\kappa (2\lambda+2\kappa -1), 
\mbox{\hspace{0.5em}} x_{\,1}^{\,1}(\lambda,\kappa)=-\tau_{\lambda}\tau_{\kappa},} \\ [0.5em]
{\scriptstyle x_{\,0}^{\,\frac{3}{2}}(\lambda,\kappa)=
(2\lambda -\tau_{\lambda})(2\kappa -\tau_{\kappa})(2\lambda+2\kappa+\tau_{\lambda}+\tau_{\kappa}-2)
(2\lambda\tau_{\kappa}+2\kappa\tau_{\lambda}-1-\tau_{\lambda}-\tau_{\kappa}+\tau_{\lambda}\tau_{\kappa}), 
\mbox{\hspace{0.5em}} x_{\,1}^{\,\frac{3}{2}}(\lambda,\kappa)=0.}  
\end{array}
\end{equation} 
Furthermore, from $\omega>1$, 
it can be seen that
\begin{equation}   \label{eq:xomegomega}
x_{\,[\omega]}^{\;\omega}(\lambda,\kappa)=0,
\end{equation}
therefore $\left(c_{0}+{d_{0}}^{2}\right)$ is factorizable, such that one can write: 
\begin{eqnarray}   \label{eq:FormofPolynPbis}
 {\cal P}^{\,\omega}(\lambda,\kappa)=\left( {d_{0}} \right)^{\tau_{\omega}} \left(c_{0}+{d_{0}}^{2}\right)  \sum_{m=0}^{m=[\omega]-1} 
x_{\,m}^{\,\omega}(\lambda,\kappa) \nonumber \\
\times   
\left(c_{0}+{d_{0}}^{2}\right)^{[\omega]-1-m}({d_{0}}^{2})^{m}, \mbox{\hspace{1em}} \forall \omega > 1. 
\end{eqnarray}

           $\bullet$ {\tt Recursion relations for polynomials} ${\cal P}^{\,\omega}(\lambda,\kappa)$:\par
\hspace*{1em} As can be seen, eqs.~ (\ref{eq:RecurrPpIter})-(\ref{eq:RecurrPpiIter})
are not straightforwardly suitable for finding general formulas.
Actually, they are functional relations rather than really usable recursion relations
over the degrees of polynomials ${\cal P}$. Then, we need to transform them in such a way that
recursive methods can really be operational.\par
For instance, for eq.~(\ref{eq:RecurrPpIter}), a first iteration
is carried out thanks to the change $\lambda \rightarrow \lambda -\frac{1}{2}$,
$\kappa \rightarrow \kappa +\frac{1}{2}$. One removes the resulting expression into eq.~(\ref{eq:RecurrPpIter}). This leads to an equation of order
$1$. This process is repeated with changes like $\lambda \rightarrow \lambda - 1$,
$\kappa \rightarrow \kappa +1$, and so on, until it stops at {\em a final order} $m$. osp(1$|$2)-triangular constraints imply that
\begin{equation}   \label{eq:EndIterp}
[\mbox{\boldmath$S$}^{\frac{p-1}{2}}\times \mbox{\boldmath$S$}^{\lambda+\kappa-\frac{p-1}{2}}]^{\lambda+\kappa-p}=0,
\end{equation}
whence
\begin{equation}   \label{eq:finalorderp}
m=2\lambda -p.
\end{equation}
For eq.~(\ref{eq:RecurrPpiIter}) the same method can be applied, and the result will be given with $\pi=p+\frac{1}{2}$. 
Use of explicit expressions of $\alpha_{\kappa}$ and $\gamma_{\kappa}$ (cf. subsects. \ref{alphacalcul}, \ref{gammacalcul})
allows one to present  the desired final results as shown below.\par

\begin{eqnarray}   \label{eq:IterablePorderp}
\lefteqn{ \frac{(2\kappa-p)!}{(2\kappa)!} {\cal P}^{\,p}(\lambda,\kappa)= 
\displaystyle{\sum_{n=0}^{n=2\lambda -p}}
\frac{(2\kappa+n-p)!}{(2\kappa+n)!}} \nonumber \\
& & 
\begin{array}{c}
\times \displaystyle \frac{(2\kappa +n +\tau_{\kappa+\frac{n}{2}})}{(2\kappa+n)(2\kappa+n+1)} \,d_{0} 
\,{\cal P}^{\,p-\frac{1}{2}}\begin{array}{c}(\lambda-\frac{(n+1)}{2},\kappa+\frac{n}{2}) \end{array} \\  [0.7em]
+ (2\lambda+2\kappa -p) \displaystyle{\sum_{n=0}^{n=2\lambda -p}}
\frac{(2\kappa+n-p)!}{(2\kappa+n)!} \\ [1.2em]
\times \displaystyle \left((2\kappa+n)(c_{0}+{d_{0}}^{2})-\frac{\tau_{\kappa+\frac{n}{2}}}{(2\kappa+n)} {d_{0}}^{2}\right)  
\,{\cal P}^{\,p-1}\begin{array}{c}(\lambda-\frac{(n+1)}{2},\kappa+\frac{(n-1)}{2}). \end{array}
\end{array} \nonumber \\ 
\end{eqnarray}
\begin{eqnarray}   \label{eq:IterablePorderpp12}
\lefteqn{ \frac{(2\kappa-p-1)!}{(2\kappa)!} {\cal P}^{\,p+\frac{1}{2}}(\lambda,\kappa)=  
(2\lambda+2\kappa -p) \displaystyle{\sum_{n=0}^{n=2\lambda -p-1}}
(-1)^{n}\frac{(2\kappa+n-p)!}{(2\kappa+n)!}
} \nonumber \\
& & 
\begin{array}{c}
\times \displaystyle \frac{(2\kappa +n +\tau_{\kappa+\frac{n}{2}})}{(2\kappa+n)(2\kappa+n+1)} \,d_{0} 
\,{\cal P}^{\,p}\begin{array}{c}(\lambda-\frac{(n+1)}{2},\kappa+\frac{n}{2}) \end{array} \\ [0.7em]
-(2\lambda+2\kappa -p) \displaystyle{\sum_{n=0}^{n=2\lambda -p-1}}
(-1)^{n}\frac{(2\kappa+n-p-1)!}{(2\kappa+n)!} \\ [1.2em]
\times \displaystyle \left((2\kappa+n)(c_{0}+{d_{0}}^{2})-\frac{\tau_{\kappa+\frac{n}{2}}}{(2\kappa+n)} {d_{0}}^{2}\right)  
\,{\cal P}^{\,p-\frac{1}{2}}\begin{array}{c}(\lambda-\frac{(n+1)}{2},\kappa+\frac{(n-1)}{2}). \end{array}
\end{array} \nonumber \\ 
\end{eqnarray}
Now, these (functional) relations are ``iterable'' over $p$ and convenient for a step analogous to step (iv) of sect.~\ref{MethodIter}.
This time, calculations are free from factors in square roots.\par \vspace{0.7em}

$\bullet$ {\tt Recursion relations for expansion coefficients $x_{\,m}^{\,\omega}(\lambda,\kappa)$} \vspace{0.7em}\par
Use of eqs. (\ref{eq:FormofPolynP}), (\ref{eq:IterablePorderp}) allows one to carry out an identification over
$\left(c_{0}+{d_{0}}^{2}\right)^{p-m}({d_{0}}^{2})^{m}$. This leads to two boundary equations and a general one.\par

{\bf - \small Boundary coefficient} of $\left(c_{0}+{d_{0}}^{2}\right)^{p}$:
\begin{eqnarray} \label{eq:Recurrxcoeffc0d02powerp}
\lefteqn{
\frac{(2\kappa-p)!}{(2\kappa)!}x_{\,0}^{\,p}(\lambda,\kappa)=  } \nonumber \\
&& (2\lambda+2\kappa-p)\displaystyle \sum_{n=0}^{n=2\lambda-p}
\frac{(2\kappa+n-p)!}{(2\kappa+n-1)!}
x_{\,0}^{\,p-1}\begin{array}{c}(\lambda-\frac{(n+1)}{2},\kappa+\frac{(n-1)}{2})\end{array}. \nonumber \\
\end{eqnarray}   

{\bf - \small Boundary coefficient} of ${({d_{0}}^{2})}^{p}$:
\begin{eqnarray}   \label{eq:Recurrxcoeffd02powerp}
\mbox{\hspace*{1em}}
\lefteqn{
\frac{(2\kappa-p)!}{(2\kappa)!}x_{\,p}^{\,p}(\lambda,\kappa)=  } \nonumber \\
&& \begin{array}{c} 
\displaystyle \sum_{n=0}^{n=2\lambda-p}
\frac{(2\kappa+n-p)!}{(2\kappa+n)!} \frac{(2\kappa+n+\tau_{\kappa+\frac{n}{2}})}{(2\kappa+n)(2\kappa+n+1)}
x_{\,p-1}^{\,p-\frac{1}{2}}\begin{array}{c}(\lambda-\frac{(n+1)}{2},\kappa+\frac{n}{2})\end{array}  \nonumber \\ [1.5em]
-(2\lambda+2\kappa-p)\displaystyle \sum_{n=0}^{n=2\lambda-p} \frac{(2\kappa+n-p)!}{(2\kappa+n)!}
\frac{\tau_{\kappa+\frac{n}{2}}}{(2\kappa+n)}
x_{\,p-1}^{\,p-1}\begin{array}{c}(\lambda-\frac{(n+1)}{2},\kappa+\frac{(n-1)}{2})\end{array}. \nonumber \\
\end{array} \nonumber \\ 
\end{eqnarray}
{\bf - \small General coefficients} of $\left(c_{0}+{d_{0}}^{2}\right)^{p-m}{({d_{0}}^{2})}^{m}$ with $1\leq m$ and $p \geq 1$:
\begin{eqnarray}   \label{eq:Recurrxcoeffgeneral}
\mbox{\hspace*{1.5em}}
\lefteqn{
\frac{(2\kappa-p)!}{(2\kappa)!}x_{\,m}^{\,p}(\lambda,\kappa)=  } \nonumber \\
&& \begin{array}{c} 
\displaystyle \sum_{n=0}^{n=2\lambda-p}
\frac{(2\kappa+n-p)!}{(2\kappa+n)!} \frac{(2\kappa+n+\tau_{\kappa+\frac{n}{2}})}{(2\kappa+n)(2\kappa+n+1)}
x_{\,m-1}^{\,p-\frac{1}{2}}\begin{array}{c}(\lambda-\frac{(n+1)}{2},\kappa+\frac{n}{2})\end{array}  \nonumber \\ [1.5em]
+ (2\lambda+2\kappa-p)\displaystyle \sum_{n=0}^{n=2\lambda-p}
\frac{(2\kappa+n-p)!}{(2\kappa+n-1)!} x_{\,m}^{\,p-1}\begin{array}{c}(\lambda-\frac{(n+1)}{2},\kappa+\frac{(n-1)}{2})\end{array} 
\nonumber \\ [1.5em]
-(2\lambda+2\kappa-p)\displaystyle \sum_{n=0}^{n=2\lambda-p} \frac{(2\kappa+n-p)!}{(2\kappa+n)!}
\frac{\tau_{\kappa+\frac{n}{2}}}{(2\kappa+n)}
x_{\,m-1}^{\,p-1}\begin{array}{c}(\lambda-\frac{(n+1)}{2},\kappa+\frac{(n-1)}{2})\end{array}. \nonumber
\end{array} \nonumber \\ 
\end{eqnarray}

\hspace*{1.5em} The same method of identification is used from eq.~(\ref{eq:IterablePorderpp12}) instead of eq.~(\ref{eq:IterablePorderp}). After a frontal simplification by $d_0$, the respective equations analogous to eqs.~(\ref{eq:Recurrxcoeffc0d02powerp})-(\ref{eq:Recurrxcoeffgeneral}) are found to be the following:

\begin{eqnarray}   \label{eq:Recurrxp12c0pd02powerp}
\mbox{\hspace*{2.5em}}
\lefteqn{
\frac{(2\kappa-p-1)!}{(2\kappa)!}x_{\,0}^{\,p+\frac{1}{2}}(\lambda,\kappa)= (2\lambda+2\kappa-p) } \nonumber \\
&& \begin{array}{c} 
\times \displaystyle \sum_{n=0}^{n=2\lambda-p-1} (-1)^{n}
\frac{(2\kappa+n-p)!}{(2\kappa+n)!} \frac{(2\kappa+n+\tau_{\kappa+\frac{n}{2}})}{(2\kappa+n)(2\kappa+n+1)}
x_{\,0}^{\,p}\begin{array}{c}(\lambda-\frac{(n+1)}{2},\kappa+\frac{n}{2})\end{array}  \nonumber \\ [1.5em]
-(2\lambda+2\kappa-p)\displaystyle \sum_{n=0}^{n=2\lambda-p-1} (-1)^{n} \frac{(2\kappa+n-p-1)!}{(2\kappa+n-1)!}
x_{\,0}^{\,p-\frac{1}{2}}\begin{array}{c}(\lambda-\frac{(n+1)}{2},\kappa+\frac{(n-1)}{2})\end{array}, \nonumber
\end{array} \nonumber \\ 
\end{eqnarray}

\begin{eqnarray}   \label{eq:Recurrxp12d02powerp}
\mbox{\hspace*{2.5em}}
\lefteqn{
\frac{(2\kappa-p-1)!}{(2\kappa)!}x_{\,p}^{\,p+\frac{1}{2}}(\lambda,\kappa)= (2\lambda+2\kappa-p) } \nonumber \\
&& \begin{array}{c} 
\times \displaystyle \sum_{n=0}^{n=2\lambda-p-1} (-1)^{n}
\frac{(2\kappa+n-p)!}{(2\kappa+n)!} \frac{(2\kappa+n+\tau_{\kappa+\frac{n}{2}})}{(2\kappa+n)(2\kappa+n+1)}
x_{\,p}^{\,p}\begin{array}{c}(\lambda-\frac{(n+1)}{2},\kappa+\frac{n}{2})\end{array}  \nonumber \\ [1.5em]
+(2\lambda+2\kappa-p)   \nonumber \\ 
\times \displaystyle \sum_{n=0}^{n=2\lambda-p-1} (-1)^{n} \frac{(2\kappa+n-p-1)!}{(2\kappa+n)!} \frac{\tau_{\kappa+\frac{n}{2}}}{(2\kappa+n)}
x_{\,p-1}^{\,p-\frac{1}{2}}\begin{array}{c}(\lambda-\frac{(n+1)}{2},\kappa+\frac{(n-1)}{2})\end{array}, \nonumber
\end{array} \nonumber \\ 
\end{eqnarray}

\begin{eqnarray}   \label{eq:Recurrxp12general}
\mbox{\hspace*{2.5em}}
\lefteqn{
\frac{(2\kappa-p-1)!}{(2\kappa)!}x_{\,m}^{\,p+\frac{1}{2}}(\lambda,\kappa)= (2\lambda+2\kappa-p) } \nonumber \\
&& \begin{array}{c} 
\times \displaystyle \sum_{n=0}^{n=2\lambda-p-1} (-1)^{n}
\frac{(2\kappa+n-p)!}{(2\kappa+n)!} \frac{(2\kappa+n+\tau_{\kappa+\frac{n}{2}})}{(2\kappa+n)(2\kappa+n+1)}
x_{\,m}^{\,p}\begin{array}{c}(\lambda-\frac{(n+1)}{2},\kappa+\frac{n}{2})\end{array}  \nonumber \\ [1.5em]
-(2\lambda+2\kappa-p) \displaystyle \sum_{n=0}^{n=2\lambda-p-1} (-1)^{n} \frac{(2\kappa+n-p-1)!}{(2\kappa+n-1)!}
x_{\,m}^{\,p-\frac{1}{2}}\begin{array}{c}(\lambda-\frac{(n+1)}{2}),\kappa+\frac{(n-1)}{2})\end{array} \nonumber \\
+(2\lambda+2\kappa-p)   \nonumber \\ 
\times \displaystyle \sum_{n=0}^{n=2\lambda-p-1} (-1)^{n} \frac{(2\kappa+n-p-1)!}{(2\kappa+n)!} \frac{\tau_{\kappa+\frac{n}{2}}}{(2\kappa+n)}
x_{\,m-1}^{\,p-\frac{1}{2}}\begin{array}{c}(\lambda-\frac{(n+1)}{2},\kappa+\frac{(n-1)}{2})\end{array}. \nonumber
\end{array} \nonumber \\ 
\end{eqnarray}

{ \tt Exact general results for some coefficients:} \par
\hspace*{1.5em} An instructive task should be done here by the reader himself, for retrieving the set given by eq.~(\ref{eq:Firstxomega})
from some recursion equations among eqs.~(\ref{eq:Recurrxcoeffc0d02powerp})-(\ref{eq:Recurrxp12general}), taken into account the initial 
value $x_{\,0}^{\,0}(\lambda,\kappa)=1$. In doing that indeed, it will be realized that our so-called ``summations of finite series", 
see step $(iv)$, sect.~(\ref{MethodIter}), actually consists only in canceling out gradually a lot of terms, until one or two remaining terms lead to a final result. In addition, it will be noticed that the use of beforehand symmetrized expressions in $(\lambda, \kappa)$ (related to $p-\frac{1}{2}, p-1$
pseudo-degrees needed in the recursion process, for instance)
{\em does not furnish the result under a straightforward symmetrical form in $(\lambda, \kappa)$}. An ultimate heuristic task remains to be carried out for achieving the calculation.\par  
\hspace*{1.5em} Nevertheless, definitive analytical formulas for boundary coefficients may be summed up, as presented below:\par
Solution of eq. (\ref{eq:Recurrxcoeffc0d02powerp}) can be found and reads:
\begin{equation}   \label{eq:Exactxp0}
x_{\,0}^{\,p}(\lambda,\kappa)=
\frac{(2\lambda)!(2\kappa)!(2\lambda+2\kappa-p)!}
{p!(2\lambda-p)!(2\kappa-p)!(2\lambda+2\kappa-2p)!}.
\end{equation}

Thanks to this result, eq.~(\ref{eq:Recurrxp12c0pd02powerp}) may be transformed, and exhibits a binomial coefficient:
\begin{eqnarray}   \label{eq:Recurrxp12d02pwerp1}
\mbox{\hspace*{2.5em}}
\lefteqn{
x_{\,0}^{\,p+\frac{1}{2}}(\lambda,\kappa)= \frac{(2\kappa)!}{(2\kappa-p-1)!}(2\lambda+2\kappa-p)} \nonumber \\
&& \times \left\{ \begin{array}{c} 
\displaystyle \frac{(2\lambda+2\kappa-1-p)!} {(2\lambda+2\kappa-1-2p)!}  \sum_{n=0}^{n=2\lambda-p-1} (-1)^{n}
{ 2\lambda-n-1\choose p} 
\frac{(2\kappa+n+\tau_{\kappa+\frac{n}{2}})}{(2\kappa+n)(2\kappa+n+1)} \nonumber \\ [1.5em]
-\displaystyle \sum_{n=0}^{n=2\lambda-p-1} (-1)^{n} \frac{(2\kappa+n-p-1)!}{(2\kappa+n-1)!}
x_{\,0}^{\,p-\frac{1}{2}}\begin{array}{c}(\lambda-\frac{(n+1)}{2},\kappa+\frac{(n-1)}{2})\end{array}. \nonumber
\end{array} \right. \nonumber \\ 
\end{eqnarray}
In spite of its appearance, this equation does not mean that $x_{\,0}^{\,p+\frac{1}{2}}(\lambda,\kappa)$ is a multiple of $(2\lambda+2\kappa-p)$. \par
\hspace*{1.5em} Another point regards the announced property (\ref{eq:xomegomega}). The clue for a rigorous proof  lies only in  
the result $x_{\,1}^{\,1}(\lambda,\kappa)=-\tau_{\lambda}\tau_{\kappa}$. The way consists in considering eq.~(\ref{eq:Recurrxcoeffd02powerp})
for $p=2$. This leads to $x_{\,2}^{\,2}(\lambda,\kappa)=0$. Then, one sets $p=2$ in eq.~(\ref{eq:Recurrxp12d02powerp}), which yields
$x_{\,2}^{\,\frac{5}{2}}(\lambda,\kappa)=0$, and so on. Reasoning by recursion gives the (rather unexpected) result (\ref{eq:xomegomega}), 
namely $x_{\,[\omega]}^{\;\omega}(\lambda,\kappa)=0, \forall \omega >1$. \par
\hspace*{1.5em}By using the decomposition shown in eq.~(\ref{eq:Formulalphakappa2}), making use of the standard recursion relation for binomial coefficients and taking into account the
possible various parities of $\kappa$ and $2\lambda-p-1$, the first summation in 
eq.~(\ref{eq:Recurrxp12d02pwerp1}) reduces to:

\begin{eqnarray}   \label{eq:sumbin}
\lefteqn{
\sum_{n=0}^{n=2\lambda-p-1} (-1)^{n}
\displaystyle { 2\lambda-n-1\choose p} 
\frac{(2\kappa+n+\tau_{\kappa+\frac{n}{2}})}{(2\kappa+n)(2\kappa+n+1)}= } \nonumber \\
& &  \begin{array}{c} \displaystyle {2\lambda \choose p} \frac{\tau_{\kappa}}{2\kappa} 
+(-1)^{2\kappa} {p \choose \tau_{\lambda+\kappa+\frac{p+1}{2}}} \displaystyle \frac{1}{(2\lambda+2\kappa-p-\tau_{\lambda+\kappa+\frac{p+1}{2}})} \nonumber \\  
+(-1)^{2\kappa}\displaystyle  \sum_{m=0}^{m=\left[\frac{ 2\lambda-p-3}{2}\right] +\tau_{\kappa}\tau_{\lambda+\frac{p+1}{2}}}
{ 2\lambda-2m-2+\tau_{k} \choose p-1} 
\frac{1}{(2\kappa+2m+1-\tau_{k})} . \end{array}\nonumber \\
\end{eqnarray}
In the same way, the second summation in eq.~(\ref{eq:Recurrxp12d02pwerp1}) can be re-written as follows:
\begin{eqnarray}   \label{eq:sumbin2}
\sum_{n=0}^{n=2\lambda-p-1} (-1)^{n} \frac{(2\kappa+n-p-1)!}{(2\kappa+n-1)!}
x_{\,0}^{\,p-\frac{1}{2}}\begin{array}{c}(\lambda-\frac{(n+1)}{2},\kappa+\frac{(n-1)}{2})\end{array}= \nonumber \\
\begin{array}{c} \displaystyle
\frac{(2\kappa-1-p)!}{(2\kappa-1)!} x_{\,0}^{\,p-\frac{1}{2}}\begin{array}{c}(\lambda-\frac{1}{2},\kappa-\frac{1}{2})\end{array} \nonumber \\ 
+\displaystyle \sum_{m=0}^{m=\left[\frac{ 2\lambda-p-3}{2}\right]}
 \frac{(2\kappa+2m+1-p)!}{(2\kappa+2m+1)!}
x_{\,0}^{\,p-\frac{1}{2}}\begin{array}{c}(\lambda-m-\frac{3}{2},\kappa+m+\frac{1}{2})\end{array} \nonumber \\ [1.5em]
-\displaystyle \sum_{m=0}^{m=\left[\frac{ 2\lambda-p-3}{2}\right] +\tau_{\lambda+\frac{p+1}{2}}}
\frac{(2\kappa+2m-p)!}{(2\kappa+2m)!}
x_{\,0}^{\,p-\frac{1}{2}}\begin{array}{c}(\lambda-m-1,\kappa+m)\end{array} .\nonumber
\end{array} \nonumber \\
\end{eqnarray}
Eqs.~(\ref{eq:sumbin})-(\ref{eq:sumbin2}) enable us to calculate step by step the analytical expression for the first 
$x_{\,0}^{\,p+\frac{1}{2}}$ coefficients up to $p=3$, for instance. 
According to the notations defined in table~(\ref{eq:NotationsParities}),  this leads to the following formulas
written below.\par
{\tt Analytical expression of the first expressions of} $x_{0}^{p+\frac{1}{2}}(\lambda,\kappa)$ {\tt coefficients}: \vspace{1em}
\[             \begin{array}{c|c|}
 p=0 \, & x_{0}^{\frac{1}{2}}(\lambda,\kappa)  \\ [0.5em] \hline 
\scriptstyle (a) & {\scriptstyle 0} \\ \hline 
\scriptstyle (b) & {\scriptstyle 2\lambda+2\kappa} \\ \hline
\scriptstyle (c) & {\scriptstyle 2\kappa} \\ \hline
\scriptstyle (d) & {\scriptstyle 2\lambda} \\ \hline
\end{array}        \]

\[             \begin{array}{c|c|}
 p=1 \, & x_{0}^{\frac{3}{2}}(\lambda,\kappa)  \\ [0.5em] \hline 
\scriptstyle (a) 
& {\scriptstyle -2\lambda \cdot 2\kappa (2\lambda+2\kappa-2)} \\ \hline
\scriptstyle (b) 
& {\scriptstyle (2\lambda-1)(2\kappa-1)(2\lambda+2\kappa)(2\lambda+2\kappa-2)} \\ \hline
\scriptstyle (c) 
& {\scriptstyle (2\lambda-1)2\kappa(2\kappa-2)(2\lambda+2\kappa-1)} \\ \hline
\scriptstyle (d) 
& {\scriptstyle 2\lambda(2\lambda-2)(2\kappa-1)(2\lambda+2\kappa-1)} \\ \hline
\end{array}        \]

\[             \begin{array}{c|c|}
 p=2 \, & x_{0}^{\frac{5}{2}}(\lambda,\kappa)  \\ [0.5em] \hline 
\scriptstyle (a) 
& {\scriptstyle -2\lambda(2\lambda-2)2\kappa(2\kappa-2) (2\lambda+2\kappa-2)(2\lambda+2\kappa-4)} \\ \hline
\scriptstyle (b) 
& {\scriptstyle  \frac{1}{2} (2\lambda-1)(2\kappa-1)(2\lambda+2\kappa-2)(2\lambda+2\kappa-4)
\bigl( (2\lambda-2)(2\kappa-2)(2\lambda+2\kappa-1)+4\bigr) } \\ \hline
\scriptstyle (c) 
& {\scriptstyle  \frac{1}{2} (2\lambda-1)(2\kappa)(2\kappa-2)(2\lambda+2\kappa-3)
\bigl( (2\lambda-2)(2\kappa-3)(2\lambda+2\kappa-2)+4\bigr) } \\ \hline
\scriptstyle (d) 
& {\scriptstyle \frac{1}{2} (2\lambda)(2\lambda-2)(2\kappa-1)(2\lambda+2\kappa-3)
\bigl( (2\lambda-3)(2\kappa-2)(2\lambda+2\kappa-2)+4\bigr) } \\ \hline
\end{array}        \]

\[             \begin{array}{c|c|}
 p=3 \, & x_{0}^{\frac{7}{2}}(\lambda,\kappa)  \\ [0.5em] \hline 
\scriptstyle (a)
& {\scriptstyle  -\frac{1}{2} (2\lambda)(2\lambda-2)2\kappa(2\kappa-2)(2\lambda+2\kappa-4)(2\lambda+2\kappa-6)
\bigl( (2\lambda-3)(2\kappa-3)(2\lambda+2\kappa-3)+4\bigr) } \\ \hline
\scriptstyle (b) 
& { \begin{array}{c} \scriptstyle \frac{1}{6} (2\lambda-1)(2\lambda-3)(2\kappa-1)(2\kappa-3)(2\lambda+2\kappa-2)(2\lambda+2\kappa-4)(2\lambda+2\kappa-6) \\ 
 \scriptstyle \times \bigl( (2\lambda-2)(2\kappa-2)(2\lambda+2\kappa-3)+12\bigr) \end{array} }\\ \hline
\scriptstyle (c) 
& { \begin{array}{c} \scriptstyle  \frac{1}{6} (2\lambda-1)(2\lambda-3)2\kappa(2\kappa-2)(2\kappa-4)(2\lambda+2\kappa-3)(2\lambda+2\kappa-5) \\ 
\scriptstyle \times \bigl( (2\lambda-2)(2\kappa-3)(2\lambda+2\kappa-4)+12\bigr) \end{array} } \\ \hline
\scriptstyle (d) 
& { \begin{array}{c} \scriptstyle \frac{1}{6} (2\lambda)(2\lambda-2)(2\lambda-4)(2\kappa-1)(2\kappa-3)(2\lambda+2\kappa-3)(2\lambda+2\kappa-5) \\ 
\scriptstyle \times \bigl( (2\lambda-3)(2\kappa-2)(2\lambda+2\kappa-4)+12\bigr) \end{array} } \\ \hline
\end{array} \]
{\em Remarks:} It can be experimented that for $p=3$ the calculation becomes rather laborious, thus discouraging us to go further
with the hope of guessing a general formula for any $p$. Our calculations show that, in a given step, relevant sums like
$\sum_{m=0} 1/(2\kappa+m)$ vanish systematically. This is welcome because these sums actually are transcendental and could be expressed as function of Riemann Zeta function $\zeta(z,q)$.

                           \subsection{Polynomials ${\cal Q}^{\,\omega}$}
From eq.~(\ref{eq:ReccurSLow}) and hypothesis (\ref{eq:ClosRelosp12second}), primary recursion relations for ${\cal Q}$ are found to be: 
\begin{eqnarray}   \label{eq:RecurrQp}
2\kappa\begin{array}{c}{\cal Q}^{\,p}(\lambda;\kappa)\end{array}=(2\kappa +p)
\begin{array}{c}{\cal Q}^{\,p}(\lambda-\frac{1}{2};\kappa-\frac{1}{2})\end{array} \nonumber \\
+(-1)^{2\lambda}\frac{(2\kappa +\tau_{\kappa})}{2\kappa+1} d_{0} \begin{array}{c}{\cal Q}^{\,p-\frac{1}{2}}(\lambda-\frac{1}{2};\kappa)\end{array}
 \nonumber \\
+2\kappa(2\kappa-2\lambda+p+1) \gamma_{\kappa+\frac{1}{2}} 
\begin{array}{c}{\cal Q}^{\,p-1}(\lambda-\frac{1}{2};\kappa+\frac{1}{2})\end{array}, 
\end{eqnarray}
\begin{eqnarray}   \label{eq:RecurrQpi}
2\kappa\begin{array}{c}{\cal Q}^{\,\pi}(\lambda;\kappa)\end{array}=\begin{array}{c}(2\kappa +\pi+\frac{1}{2})
{\cal Q}^{\,\pi}(\lambda-\frac{1}{2};\kappa-\frac{1}{2})\end{array} \nonumber \\
-(-1)^{2\lambda}\begin{array}{c}(2\kappa+\pi+\frac{1}{2})(2\kappa-2\lambda+\pi+\frac{1}{2})\end{array}\frac{(2\kappa +\tau_{\kappa})}{2\kappa+1} 
d_{0} \begin{array}{c}{\cal Q}^{\,\pi-\frac{1}{2}}(\lambda-\frac{1}{2};\kappa)\end{array} \nonumber \\
+\begin{array}{c}2\kappa(2\kappa-2\lambda+\pi+\frac{1}{2}) \gamma_{\kappa+\frac{1}{2}}
{\cal Q}^{\,\pi-1}(\lambda-\frac{1}{2};\kappa+\frac{1}{2})\end{array}. \nonumber \\
\end{eqnarray}

                               \renewcommand{\theequation}{C.\arabic{equation}}
                               \setcounter{equation}{0}

                              \section*{Appendix C \\ [1em] Updated analysis of the attempt of Zeng \cite{Zeng.4} and his proposition for an osp(1$|$2)-triangle sum rule}\label{ZengSumRule}
\hspace*{1.5em}In order to clearly summarize the reasoning followed by Zeng, we need to proceed to a transformation of several parameters which were used in his paper: notations not in common use (osp(1$|$2) C-G coefficients), isoscalar factors (non parity-independent),
osp(1$|$2) Racah coefficients (with eight different analytical forms) and use of su(2) Racah coefficients. Thus, we will work with
parity-independent scalar factors, $6$-$j^{S}$ symbols \cite{L.B.Nuovo.I}, and usual su(2) $6$-$j$ symbols, as shown below.\par
$\bullet$ The starting point is the su(2) $\bigtriangledown$-sum rule:
\begin{eqnarray}   \label{eq:ZSumRulesu2}
\lefteqn{ 
\sum_{L_3}\bigtriangledown(l_1 l_2 L_3) \bigtriangledown(L_1 L_2 L_3)  
\left\{ \begin{array}{ccc} L_1 & l_2 & l_3 \\ l_1 & L_2 & L_3 \end{array} \right\} } \nonumber \\
&& = \frac{(-1)^{L_1+l_1+L_2+l_2}}{(2l_3+1)} \bigtriangledown(L_1 l_2 l_3) \bigtriangledown(l_1 L_2 l_3). 
\end{eqnarray}
\hspace*{1.5em}- An intermediate step, in our re-formulation, consists in the use of pseudo-orthogonality relations related to scalar factors with
a summation over $l_1, l_2$ to make the most of eq.~(8.6) of ref.~\cite{L.B.Nuovo.I}. This leads to a formula like that:
\begin{eqnarray}   \label{eq:6jSscalarfactors}
\sum_{J_{3}} \frac{(-1)^{(phases)}}{(2L_3+1)} \left\{ \begin{array}{ccc} J_1 & j_2 & j_3 \\ j_1 & J_2 & J_3 \end{array} \right\}^{\!_S}
\left[ \begin{array}{ccc} J_1 & J_2 & J_3 \\ L_1 & L_2 & L_3 \end{array} \right]
\left[ \begin{array}{ccc} j_1 & j_2 & J_3 \\ l_1 & l_2 & L_3 \end{array} \right] = \nonumber \\
\left( \sum_{l_3} \right) \frac{(-1)^{(phases)}}{(2l_3+1)}
\left\{ \begin{array}{ccc} L_1 & l_2 & l_3 \\ l_1 & L_2 & L_3 \end{array} \right\}
\left[ \begin{array}{ccc} J_1 & j_2 & j_3 \\ L_1 & l_2 & l_3 \end{array} \right]
\left[ \begin{array}{ccc} j_1 & J_2 & j_3 \\ l_1 & L_2 & l_3 \end{array} \right].  
\end{eqnarray}
Actually, there is no summation over $l_3$, that's why we have written $\left( \sum_{l_3} \right)$ with brackets. 
Indeed, if $l_1,L_1,l_2,L_2$ are fixed, then $l_3$ gets a single value, namely
\begin{equation}   \label{eq:Fixedl3}
l_3=j_3 - \begin{array}{c}\frac{1}{2}\end{array} \tau_{j_3+L_1+l_2} = j_3 - \begin{array}{c}\frac{1}{2}\end{array} \tau_{j_3+l_1+L_2}.
\end{equation}
Definition and properties of $\tau$ have been previously given, see eqs.~(\ref{eq:taudef}), (\ref{eq:tauproperties}).

$\bullet$ Next, the principle of Zeng, at this step, is to operate both sides of eq.~(\ref{eq:6jSscalarfactors}) by
${\displaystyle \sum_{L_3}} \bigtriangledown(l_1 l_2 L_3) \bigtriangledown(L_1 L_2 L_3)$ in order to make use of the 
su(2) $\bigtriangledown$-sum rule.\par
\hspace*{1.5em}- With the aid of our study on phase factors for obtaining explicit analytical formulas
of $6$-$j^{S}$ symbols involving functions of scalar factor moduli, see sect.~{\bf 8} in ref.~\cite{L.B.Nuovo.I}, one sees that an updated version of his formulation can be written in the following way:
\begin{eqnarray}   \label{eq:osp12ZengSrule}
\sumhalf{J_{3}}{}  \frac{(-1)^{\Psi_{J_{3}}}}{(2L_3+1)}
(-1)^{[j_1+j_2+J_3]+[J_1+J_2+J_3]+2J_3} 
\left\{ \begin{array}{ccc} J_1 & j_2 & j_3 \\ j_1 & J_2 & J_3 \end{array} \right\}^{\!_S}  \nonumber \\
\times \left| \left[ \begin{array}{ccc} J_1 & J_2 & J_3 \\ L_1 & L_2 & L_3 \end{array} \right] \right|
\bigtriangledown(L_1 L_2 L_3)
\left| \left[ \begin{array}{ccc} j_1 & j_2 & J_3 \\ l_1 & l_2 & L_3 \end{array} \right] \right|
\bigtriangledown(l_1 l_2 L_3)
= \nonumber \\
\frac{(-1)^{\varphi_{j_3}}}{(2l_3+1)}
\left| \left[ \begin{array}{ccc} J_1 & j_2 & j_3 \\ L_1 & l_2 & l_3 \end{array} \right] \right|
\bigtriangledown(L_1 l_2 l_3)
\left| \left[ \begin{array}{ccc} j_1 & J_2 & j_3 \\ l_1 & L_2 & l_3 \end{array} \right] \right|
\bigtriangledown(l_1 L_2 l_3)
.
\end{eqnarray}
Here also, given $J_3$, $L_3$ is fixed according to:
\begin{equation}   \label{eq:FixedL3}
L_3=J_3 - \begin{array}{c}\frac{1}{2}\end{array} \tau_{J_3+l_1+l_2} = J_3 - \begin{array}{c}\frac{1}{2}\end{array} \tau_{J_3+L_1+L_2}.
\end{equation}
Our phases are given by:
\begin{eqnarray}   \label{eq:PsiJ3}
\lefteqn{
(-1)^{\Psi_{J_{3}}}=(-1)^{4(j_1J_1+j_2J_2+j_3J_3) + 4(j_1+j_2+J_3)j_3} }  \nonumber \\
& & \times (-1)^{8(J_1+J_2+J_3)(J_1 -L_1)(J_2 -L_2) 
+8(j_1+j_2+J_3)((j_1 -l_1)(j_2 -l_2)+(l_1+L_2)(l_2+L_1))}, \nonumber \\
\end{eqnarray}
\begin{eqnarray}   \label{eq:phij3}
\lefteqn{
(-1)^{\varphi_{j_3}}=(-1)^{4(J_1 +j_2 +j_3)(j_1 -l_1) + 4(j_1+J_2+j_3)(j_2 -l_2)} } \nonumber \\ 
& & \times
(-1)^{8(J_1 +j_2 +j_3)(j_2 -l_2)(J_1 -L_1)+8(j_1+J_2+j_3)(j_1 - l_1)(J_2 -L_2)}.
\end{eqnarray}
$\bullet$ A relation {\em like} (\ref{eq:osp12ZengSrule}), just as it stands, was called by Zeng \cite{Zeng.4} ``the triangle sum rule of osp(1$|$2)".\par
\hspace*{1.5em}- The $\tau$-parity of $j_1+j_2+J_1+J_2$, integral or half-integral, is an important parameter which offers a ``free"
choice of $L_1, L_2, l_1, l_2$, {\em i.e.} $L_1=J_1$ or $L_1=J_1-\frac{1}{2}$ {\em etc}. That leads to different aspects of the analytical form taken by eq.~(\ref{eq:osp12ZengSrule}), thus depending only on $J_1, J_2, J_3, j_1, j_2, j_3$.
However, even by using our factorization of scalar factors \cite{L.B.Nuovo.I}, namely
\begin{equation}   \label{eq:ScalarFactor}
\left| \left[ \begin{array}{ccc} j_1 & j_2 & j_3 \\
l_1 & l_2 & l_3 \end{array} \right]  \right| = 
\left\{\begin{array}{l} \bigtriangledown(l_1 l_2 l_3)\bigtriangleup^{S}(j_1 j_2 j_3)
\mbox{\hspace{2em}\scriptsize{ $j_1+j_2+j_3$ integral}}\\ 
\bigtriangleup(l_1 l_2 l_3)\bigtriangledown^{S}(j_1 j_2 j_3)
 \mbox{\hspace{2em} \scriptsize{$j_1+j_2+j_3$ half-integral}}\end{array} \right. ,
\end{equation}
it can be seen that the result seems somewhat far from the concept of a veritable triangle sum rule for osp(1$|$2), such as developed in this paper.

\end{flushleft}
\end{document}